\documentclass{statsoc}
\usepackage[a4paper]{geometry}
\usepackage{graphicx}
\usepackage{multirow}
\usepackage{multicol}
\usepackage{amsmath}
\usepackage{natbib}
\usepackage{amsfonts}

\usepackage{bbm}

\usepackage{float}
\usepackage[caption = false]{subfig}
\title[Missing data patterns in runners' careers: do they matter?]{Missing data patterns in runners' careers: do they matter?
}
\author[Mattia Stival {\it et al.}]{Mattia Stival}
\address{Department of Statistical Sciences, University of Padova,
Padova,
Italy.}
\email{mattia.stival@unipd.it}
\author[Mattia Stival {\it et al.}]{Mauro Bernardi}
\address{Department of Statistical Sciences, University of Padova,
Padova,
Italy.}
\author[Mattia Stival {\it et al.}]{Manuela Cattelan}
\address{Department of Statistical Sciences, University of Padova,
Padova,
Italy.}
\author[Mattia Stival {\it et al.}]{Petros Dellaportas}
\address{Department of Statistical Science, University College London, London, Great Britain.\\
 Department of Statistics, Athens University of Economics and Business, Athens, Greece. \\
    The Alan Turing Institute, London, Great Britain.}
\begin{document}
\begin{abstract}
%
Predicting the future performance of young runners is an important research issue in experimental sports science and performance analysis. We analyse a data set with annual seasonal best performances of male middle distance runners for a period of 14 years and provide a modelling framework that accounts for both the fact that each runner has typically run in three distance events  (800, 1500 and 5000 meters) and the presence of periods of no running activities. We propose a latent class matrix-variate state space model and we empirically demonstrate that accounting for missing data patterns in runners' careers improves the out of sample prediction of their performances over time.  In particular, we demonstrate that for this analysis, the missing data patterns provide valuable information for the prediction of runner's performance.
\end{abstract}
\section{Introduction} 
Planning the future career of young runners is a relevant aspect of the work of coaches, whose role is to guide them during training so that they can perform at their best in competitions.
Identifying runner's capabilities and future possibilities is important for multiple reasons.  
It allows the training load to be appropriately allocated over the years,  for improving their performances and reducing their risk of injuries. 
Good planning, along with support during injuries, has been identified as one of the relevant factors that help avoiding drop-outs of runners \citep{dropout_iaaf}.
Moreover, good planning is important also from a psychological and emotional point of view, as it allows runners to strive for achievable goals and collect successes over the years.
Pleasant emotions, including satisfaction, have been associated with positive outcomes in, e.g. mental health, performance and engagement
\citep[][]{CECE2019128}. 
In this context, the identification of possible careers for an runner, in terms of observed personal  performance trajectories over time, is of paramount importance. 
For example, identifying the period in which runners reach their peaks can help prepare them for the most important events in their career. 
Similarly, the knowledge of the expected progress of different runners over the years provides an indication of whether the training process has been carried out correctly. The analysis of runner's trajectory is carried out in various sports.
\cite{leroy2018functional} have studied young swimmers’ progression using a functional clustering approach, while
 \cite{boccia_career_2017}  focused on individual careers of Italian long and high jumpers to figure out which characteristics of young runners are predictive of good-level results during their careers. \newline
\indent
We focus on the analysis of performances of Italian male middle distance runners, born in $1988$, in a period ranging from $2006$ to $2019$. 
Previous studies on middle distance runners are few or limited to samples with a small number of runners  \citep[see, e.g.,][]{middle_distances}. 
We use a combination of latent class and matrix-variate state space models. Latent class models for time dependent data have been extensively studied in the literature \citep[see, among others,][]{fruhwirth_schnatter_panel_2011, maharaj2019time, BartolucciMurphy}.
They allow to capture the heterogeneity in the careers of runners, thereby describing various possible observable scenarios.
Combining them with state space models offers additional advantages, including the possibility of building models for multivariate time series in an intuitive manner as well as the possibility of leveraging well-known tools for inference, including the treatment of missing data \citep{durbinkoopman}.
Unlike other types of runners and sports, middle distance runners have the major feature of competing in different distances, i.e. in the $800$, $1500$ and $5000$ meters distances, as well as in other spurious ones (i.e. the mile, $3000$ meters, etc.).
The choice of discipline in which to compete is subjective and typically associated with personal attitudes  \citep{mooses2013anthropometric}.
A runner capable of developing greater speed and power typically competes in shorter distances, with respect to those with greater endurance who compete in longer distances. 
As a consequence, observations in different distances are available for each runner over time, but the absence of a particular discipline can be informative on the runner's attitude. 
Beyond the variability among subjects related to the type of discipline performed, there is also variability in the developing of runners' careers related to both their abilities and histories. 
If a runner begins their career late in life, it is less likely that they will reach high levels; similarly, runners with unsatisfactory careers are likely to end their careers earlier, with respect to those satisfied with their performances \citep{hernandez2011age}. 
These aspects are related to drop-in and drop-out phenomena, defined as the events where runners enter and exit the observed sample, respectively. 
\newline
\indent
A key interesting and important question that naturally emerges is whether the absence of data is really associated with observed performances.  We attempt to shed light in this question by proposing a matrix state space model model in which multivariate time series are clustered together on the basis of their observed trend.
Matrix-variate state space models have found application in finance and engineering in past years \citep[see, e.g.,][]{choukroun2006kalman,wangwest2009}, but have recently gained additional interest in the statistical literature to analyze problems in which observations over time are matrices \citep[see, e.g.,][]{Hsu_MAR_spatio2021, Chen_constr_factor2020, chen_MAR2021}.
In this part of model specification, clustering is achieved via a latent selection matrix which is involved in the measurement equation. 
We propose to include temporal dynamics that aim to describe missing data patterns using two different processes. 
First, runner's personal history is described by a three state process, which describes their entry (drop-in) and exit (drop-out) from the sample.
Second, different propensities to compete in different distances are considered to describe runner's personal attitude. 
The probabilities of both the processes are assumed to be dependent on the latent classes stored in the selection matrix previously mentioned, allowing to consider the possible relation of missing data patterns with the observed performances.
In this way, clustering is not only achieved on the basis of runners' performances, but the presence and absence of data is considered informative as well.  
Works on latent class and clustering models for longitudinal data where the presence of missing values may be informative on the latent structure behind the data are few or limited to domain-specific works \citep[see, e.g.,][]{BartolucciMurphy, MIKALSEN2018569}.
\newline
\indent
Since the seminal work on missing values by \cite{rubin76}, researchers have wondered if and when it is possible to ignore the presence of missing values in their datasets. 
In this work, we consider this problem in a pure predictive framework, in which missing values will be considered as informative if having information on their presence and distribution over time helps in predicting runners performances over the years.  
If so, one could think to a causal relationship, in a Granger sense, between missing values and observed performances. 
Although coherent with our model's construction, we avoid the use of definition of informativeness of missing data in a causal sense. 
Indeed, while correlation between missing data and performances is typically expected in sports performance analysis, direct cause-effect relationships between them and their directions are not clearly defined in the sports science literature.
\newline
\indent
The  rest of the paper is organized as follows: Section \ref{subsec:data} presents a new publicly available dataset on middle distance runners; Section \ref{sec:model} describes the proposed model; Section \ref{sec:prior_specification} discusses the likelihood and the prior specification; Section \ref{sec:posterior_inference} presents the evaluation strategy of the model; Section \ref{sec:examples} shows the results with the real data. Additional details on the data, the algorithms, and the results are reported in the supplementary material accompanying the paper.
\section{Data and exploratory analysis}
\label{subsec:data}
\begin{figure}
    \centering
    \includegraphics[scale=0.35]{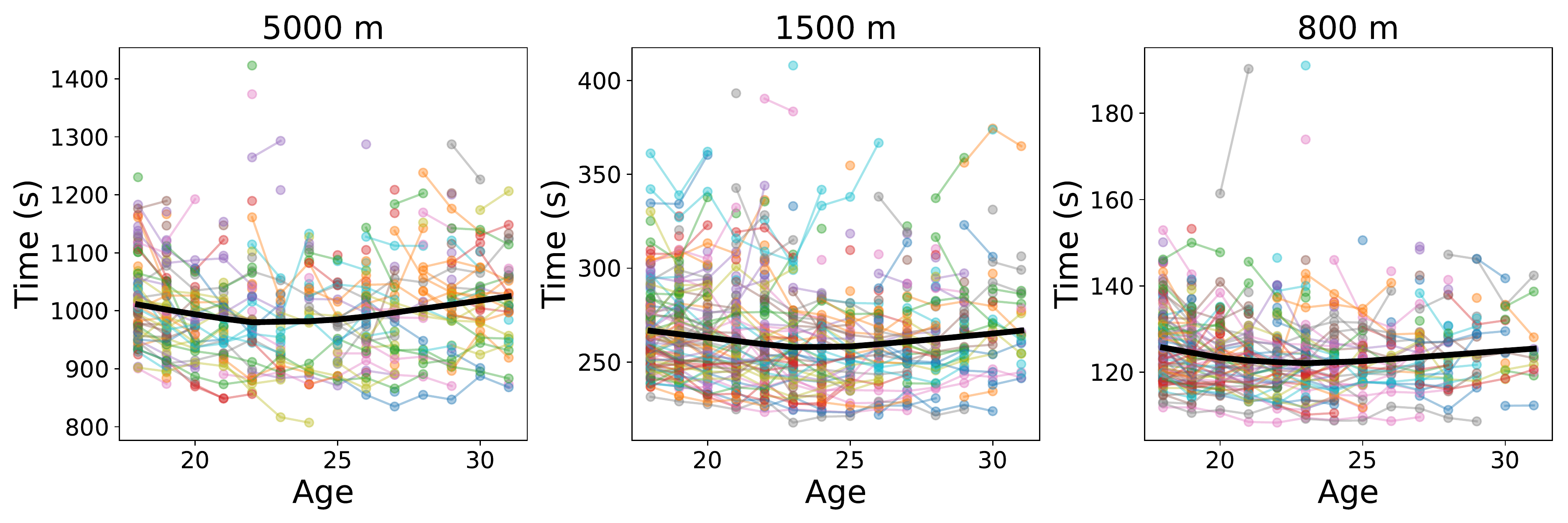}
    \caption{Seasonal best performances of $369$ Italian male middle distance runners in $5000$, $1500$, and $800$ meters distances. 
    Points represent the observed performances. 
    Lines connect the performances in consecutive years of the same runner. The black lines are obtained using local regression.}
    \label{fig:collected_data}
\end{figure}

Our data refer to annual seasonal best performances of male Italian runners, born in $1988$, on $5000$, $1500$ and $800$ meters distances in a period between $2006$ and $2019$. 
They were collected from the annual rankings accessible on the website of the Italian athletics federation (www.fidal.it), which stores results and rankings in competitions since $2005$. 
All runners with at least two observations were selected and the data are illustrated in Figure \ref{fig:collected_data} with the help of a local regression fit that allows us to perceive 
a U-shaped curve that describes the distribution of sample trajectories across ages.
U-shapes are typically observed in the evolution of runners careers \citep{performance_quadratic}. 
However, their shape can be biased by the presence of missing data, related for example to early exit or late entry in the sample.

\begin{figure}
    \centering
    \includegraphics[scale=0.5]{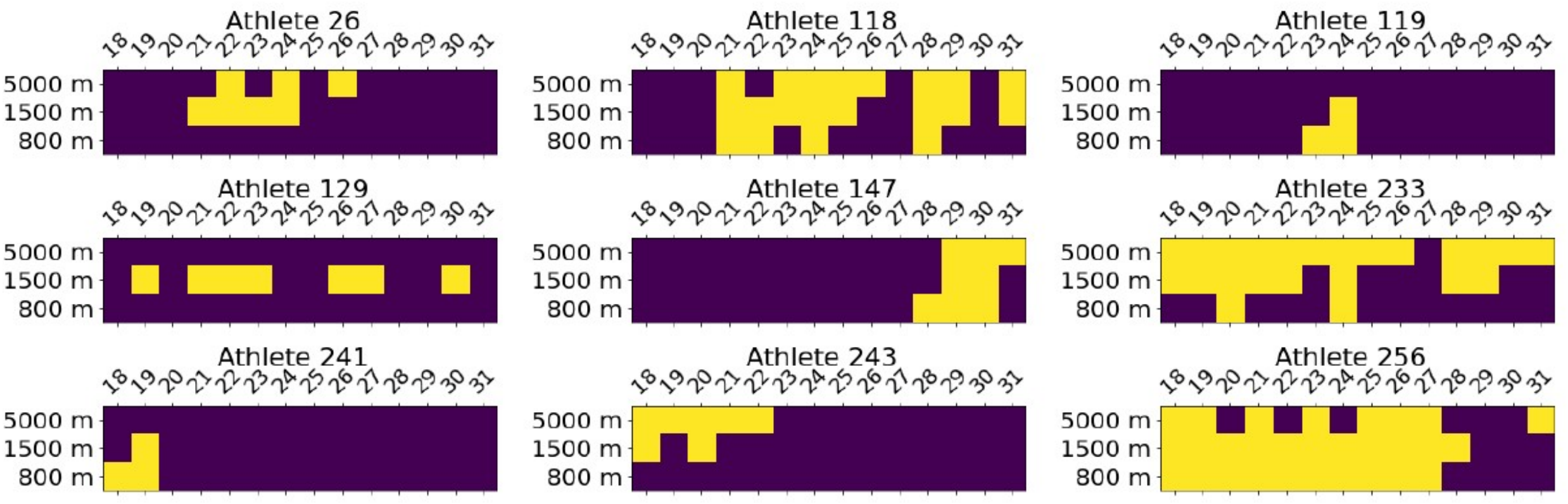}
\caption{Missing data patterns for nine runners describing their actual participation in different distances across their ages, shown on top. Yellow squares indicate that the performance is present, blue squares the absence of the observations.}
    \label{fig:missing_data_patterns}
\end{figure}

Indeed, unlike other type of data, the presence of missing values is predominant in the careers of these runners. 
Out of $15498$ observable seasonal best performances of $Q=369$ runners in $P=3$ distances and $T = 14$ years, only $2411$ seasonal best results are observed. 
A missing value is observed for one runner if the runner does not conclude (and, hence, record) any official competition in a specific discipline during one year.
The reasons for not observing any performance can be multiple. 
The runner can be \emph{not in career} in a specific year, and hence no races are performed during that year.
Alternatively, a runner in career can decide not to compete in a specific discipline for several reasons, such as lack of preparation, attitude, technical choices, etc.
To understand how missing data patterns differ between runners, Figure \ref{fig:missing_data_patterns} shows the observed patterns of missing data for nine distinct runners present in the sample. 
The patterns shown differ in various features. 
Some runners have few observations, such as runners $241$ and $119$.
Other runners are characterized by careers with many observed performances, such as runners $129$ and $256$. 
These two runners are interesting because they differ in the type of discipline they run: while runner $129$ competes only in the $1500$ meters discipline, runner $256$ competes in all the distances, recording a different number of observations for each distance.
The observed differences are typically associated both with technical choices, but also with different attitudes of the runners, leading them to compete in races of different length, according for example to their endurance and speed abilities.
We define the drop-in and drop-out as the runner age in which the  first observation in at least one discipline is present and the age after the last performance is observed, respectively.
Naturally, the careers of the runners differ both in length and the age they start racing. 
Based on this definition, runner $129$ drops-in at age $19$ and drops-out at age $31$ and runner $233$ drops-in at age $18$ and has not dropped-out in the period under examination. 

\begin{figure}
    \centering
    \subfloat[]{\includegraphics[scale=0.35]{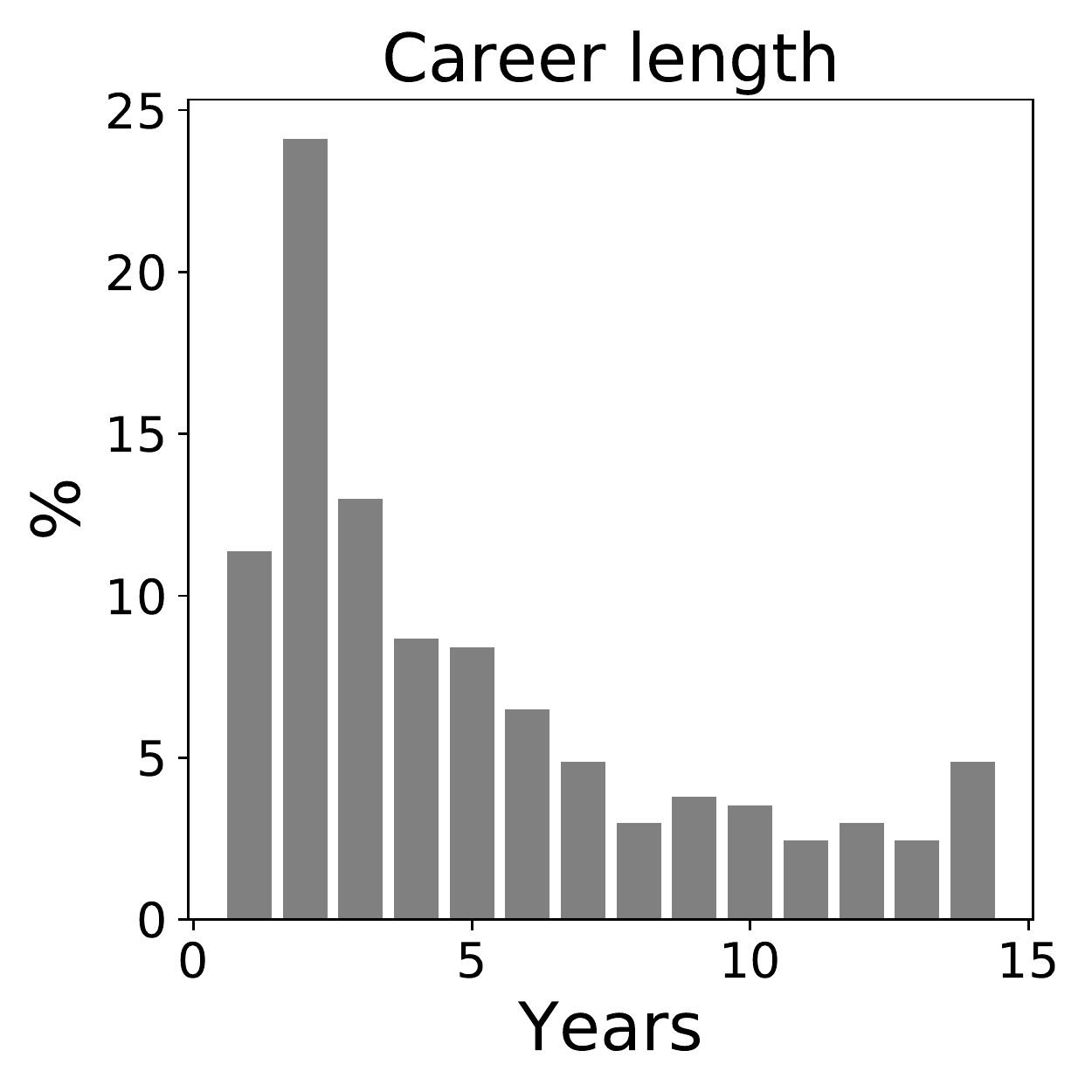}}
    \subfloat[]{\includegraphics[scale=0.35]{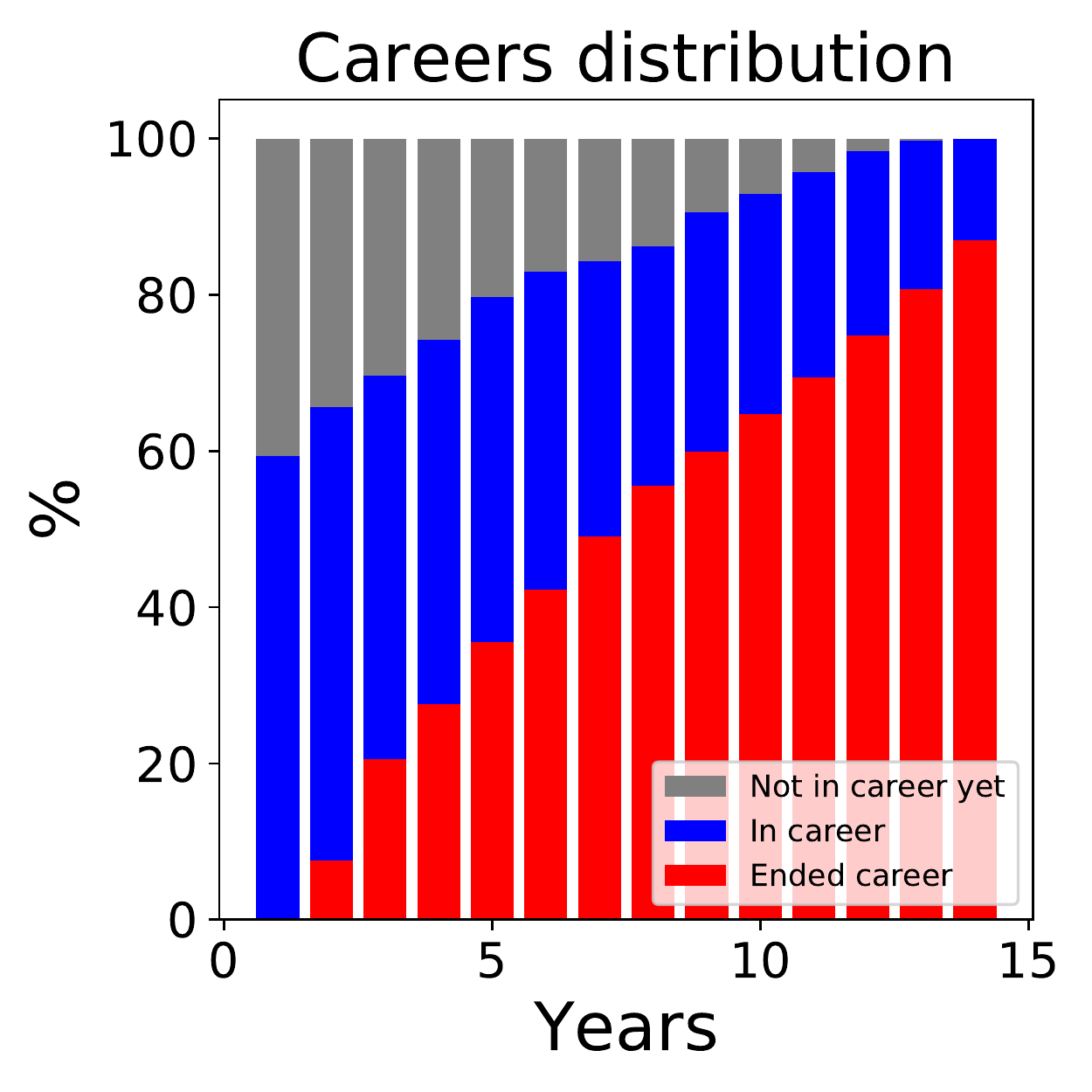}}
    \subfloat[]{\includegraphics[scale=0.35]{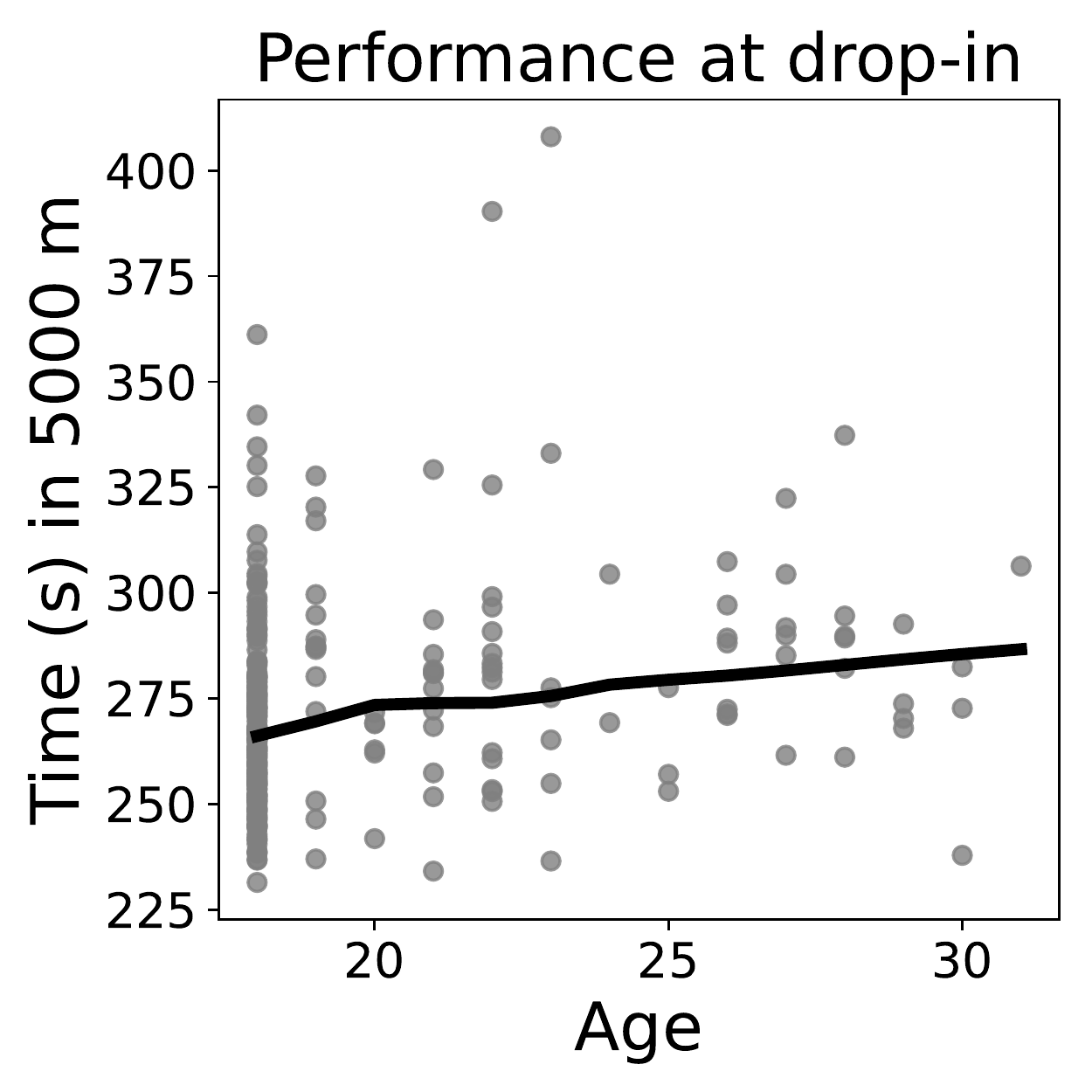}}       \\ \subfloat[]{\includegraphics[scale=0.35]{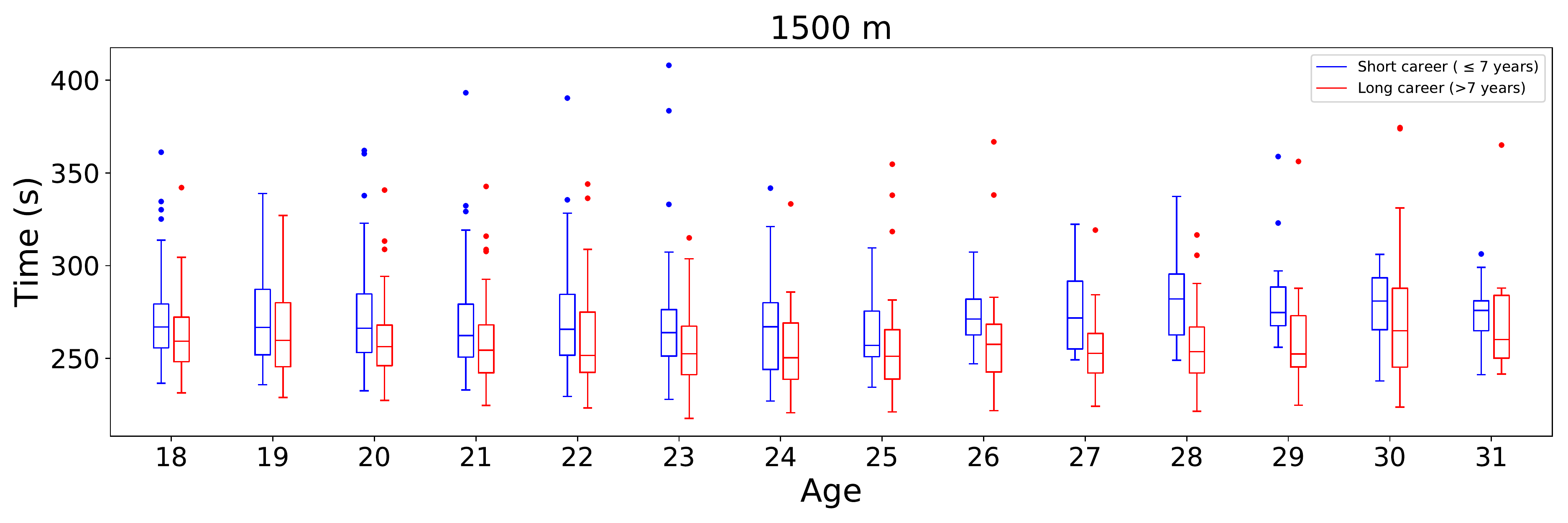}}
    \caption{
        Panel (a) shows the distribution of runners career length, panel (b) the percentage of runners that are in career (or not) in the different ages considered, and panel (c) the performances at drop-in in $5000$ meters.
        Panel (d) shows the distributions of seasonal best performances over ages in $1500$ meters discipline, grouped by the career length of the runners. Blue boxplots represent the performances of runners with a career shorter than $7$ years  (included),  red boxplots the performances of runners with a career  longer than $7$ years.
    }
    \label{fig:two_classes}
\end{figure}

The empirical distribution of runners careers' length, shown in panel (a) of Figure \ref{fig:two_classes}, is right skewed, with an average length of $5.04$ years. The increased observation at year $14$ is due to data censoring.
Panel (b) illustrates that around $60\%$ of runners in the sample competes when they are $18$ years old, but about $20\%$ of them have already left the competition at the age of $20$.
A visual exploration that indicates whether these aspects are effectively associated with observed performances are presented in panel (c) that depicts that performance at drop-in seems to be worse (greater times) if the runners start competing late in life and in panel (d) that shows the distributions of the observed performances, distinguishing between runners with careers longer or shorter than $7$ years.
In our data runners with longer careers perform typically better than those with shorter careers. 
The reason for this behavior can be either because runners with unsatisfactory career leave the competitions earlier, or because competing (and, hence, training for it) for a long period is a prerequisite for improving. We refer to the supplementary material for similar plots with other distances. 
\section{The model}
\label{sec:model}
\subsection{Clustering longitudinal data with matrix state space model}
\noindent
Let the scalar element $y_{pq,t}$ denote the observation of the performance in discipline $p$ for runner $q$ during year $t$, for $p=1,\ldots,P$, $q=1,\ldots,Q$,  and $t=1,\ldots, T$.
To facilitate the exposition, in this Section we assume that the complete set of observations is  available in the sense that  runners participate in all $P$ distances during the years and that no drop-ins or drop-outs are observed.
We assume that runners are divided into $G$ different unobserved groups according to the evolutionary trajectories during their careers. 
Suppose that runner $q$ belongs to group $g$, and that their observations over time are described by the following dynamic linear model
\begin{align}
\label{meas_clust_scalar}
    y_{pq,t}  & = \mathbf{z}_{p}^\top     \boldsymbol{\alpha}_{p,t}^{(g)} + \varepsilon_{pq,t}, 
    \\
\label{state_clust_scalar}
    \boldsymbol{\alpha}_{p,t+1}^{(g)} & = \boldsymbol{T}_p     \boldsymbol{\alpha}_{p,t}^{(g)} + \boldsymbol{\xi}_{p,t}^{(g)}, 
\end{align}
in which $\boldsymbol{\alpha}_{p,1}^{(g)} \sim \text{N}_{F_p}(\widehat{\boldsymbol{\alpha}}_{p,1\vert0}^{(g)}, \boldsymbol{P}_{p, 1\vert0}^{(g)})$, for $p=1,\ldots,P$,  $t=1,\ldots, T$, and $\widehat{\boldsymbol{\alpha}}_{p,1\vert0}^{(g)}$, $\boldsymbol{P}_{p, 1\vert0}^{(g)}$ are fixed mean and variance for the initial state, respectively. The row vector $\mathbf{z}_{p}^\top$, which has a known structure, links the observation  $y_{pq,t}$ to the column vector $\boldsymbol{\alpha}_{p,t}^{(g)}$, which describes the group-specific dynamics of the $p$--th discipline for all the runners that belong to group $g$.  
These dynamics are determined by the state transition equation that describes a first-order autoregressive process with transition matrix $\boldsymbol{T}_p$, which is discipline-specific, known, and shared across all the groups.  
In this way, for a generic discipline $p$, we require that the latent states of the different groups are different from each other, but are characterized by the same Markovian dependence induced by $\boldsymbol{T}_p$.
Moreover, this dependence is not required to be common across different distances, as $\boldsymbol{T}_p$ may differ from $\boldsymbol{T}_{p^\prime}$ for any $p\neq p^\prime$.
The error terms $\varepsilon_{pq,1},\ldots, \varepsilon_{pq,T}$ are assumed to be Gaussian with zero-mean and variances that can be discipline and subject-specific. They are assumed to be serially independent and independent of both the states $\boldsymbol{\alpha}_{p,1}^{(g)}, \ldots,\boldsymbol{\alpha}_{p,T}^{(g)}$ and the disturbances $\boldsymbol{\xi}_{p,1}^{(g)}, \ldots,\boldsymbol{\xi}_{p,T}^{(g)}$, whose covariance is $\boldsymbol{\Psi}_{pg}$, for $p = 1,\ldots, P$ and $g=1,\ldots,G$. 

Let $\mathbf{y}_{\cdot q, t} = (y_{1q,t}, \ldots, y_{Pq,t})^\top$, $\boldsymbol{\alpha}_{t}^{(g)} = (\boldsymbol{\alpha}_{1,t}^{(g)\top},\ldots,\boldsymbol{\alpha}_{P, t}^{(g)\top})^\top$, 
 $\boldsymbol{\varepsilon}_{\cdot q, t} = (\varepsilon_{1q,t}, \ldots, \varepsilon_{Pq,t})^\top$, $\boldsymbol{\xi}_{t}^{(g)} = (\boldsymbol{\xi}_{1,t}^{(g)\top},\ldots,\boldsymbol{\xi}_{P, t}^{(g)\top})^\top$, $\widehat{\boldsymbol{\alpha}}_{1 \vert 0}^{(g)} = (\widehat{\boldsymbol{\alpha}}_{1,{1 \vert 0}}^{(g)\top},\ldots,\widehat{\boldsymbol{\alpha}}_{P, {1 \vert 0}}^{(g)\top})^\top$, and $F = \sum_{p=1}^{P}F_p$. Moreover, let
 $\boldsymbol{T} = \text{blkdiag}(\boldsymbol{T}_1, \ldots,\boldsymbol{T}_P)$, $\boldsymbol{P}_{1\vert0}^{(g)} = \text{blkdiag}(\boldsymbol{P}_{1,1\vert 0}^{(g)}, \ldots,\boldsymbol{P}_{P,1\vert 0}^{(g)})$, as well as the covariance matrix $\boldsymbol{P}_{1\vert0} = \text{blkdiag}(\boldsymbol{P}_{1\vert0}^{(1)},\ldots, \boldsymbol{P}_{1\vert0}^{(G)})$ where $\text{blkdiag}(\boldsymbol{X}_a,\ldots,\boldsymbol{X}_z)$ is the block-diagonal operator, creating a block-diagonal matrix with arguments $\boldsymbol{X}_a,\ldots, \boldsymbol{X}_z$ stacked in the main diagonal. Finally, let  $\boldsymbol{Z}$ be the $P \times F$ matrix storing, in its $p$--th row, the row-vector $\mathbf{z}_p^\top$ starting from column $1-F_p+\sum_{j=1}^{p} F_j$, and zeros otherwise, and define also the following matrices:
\begin{equation*}
    \boldsymbol{Y}_t = \begin{bmatrix}     \mathbf{y}_{\cdot 1,t} & \ldots &     \mathbf{y}_{\cdot Q,t} \end{bmatrix}, \quad  \boldsymbol{A}_t = \begin{bmatrix}\boldsymbol{\alpha}_t^{(1)}&\ldots &\boldsymbol{\alpha}_t^{(G)}\end{bmatrix}, \quad     \boldsymbol{S}^\top = \begin{bmatrix} \mathbf{s}_{1\cdot}^\top & \ldots & \mathbf{s}_{Q\cdot}^\top \end{bmatrix},
\end{equation*}
\begin{equation*}
\boldsymbol{E}_t = \begin{bmatrix} \boldsymbol{\varepsilon}_{\cdot 1,t} & \ldots & \boldsymbol{\varepsilon}_{\cdot Q,t} \end{bmatrix}, \quad \boldsymbol{\Xi}_t = \begin{bmatrix}\boldsymbol{\xi}_{t}^{(1)} & \ldots & \boldsymbol{\xi}_{t}^{(G)}\end{bmatrix},\quad \widehat{\boldsymbol{A}}_{1\vert0} = \begin{bmatrix} \widehat{\boldsymbol{\alpha}}_{1 \vert 0}^{(1)}&\ldots &\widehat{\boldsymbol{\alpha}}_{1 \vert 0}^{(G)}\end{bmatrix},
\end{equation*}
 where $\mathbf{s}_{q\cdot}^\top = (\mathbbm{1}(S_{q}=1),\ldots,\mathbbm{1}(S_{q}=G))^\top$ is an allocation vector such that  $\mathbbm{1}(S_q=g) =1$ if runner $q$ belongs to group $g$, and $0$ otherwise.
Leveraging the previous notation, the model admits a matrix-variate state space representation, in which
\begin{align}
\label{clust_meas_matr}
    \boldsymbol{Y}_t & = \boldsymbol{Z}\boldsymbol{A}_t \boldsymbol{S}^\top + \boldsymbol{E}_t, \quad \boldsymbol{E}_t \sim \text{MN}_{P,Q}(\mathbf{0}, \boldsymbol{\Sigma}^C \otimes \boldsymbol{\Sigma}^R),\\
\label{clust_state_matr}
    \boldsymbol{A}_{t+1} & = \boldsymbol{T} \boldsymbol{A}_t  + \boldsymbol{\Xi}_t, \quad \quad  \boldsymbol{\Xi}_t \sim \text{MN}_{F,G}(\mathbf{0}, \boldsymbol{\Psi}^C \otimes \boldsymbol{\Psi}^R),
\end{align} 
with $\boldsymbol{A}_1 \sim \text{MN}_{F,G}(\widehat{\boldsymbol{A}}_{1\vert0}, \boldsymbol{P}_{1\vert 0})$. 
The matrix $\boldsymbol{S}$  in Equation \eqref{clust_meas_matr} is a selection matrix, with the role of selecting, for each runner, the columns of states associated with the group the runner belongs to, and silencing the others. 
The matrices of errors and disturbances are assumed to follow a matrix-variate Normal distribution with covariance matrix decomposed by a Kronecker product \citep{guptanagar}, which is a typical assumption in models for matrix-variate time series \citep[see, e.g.,][]{wangwest2009, Chen_constr_factor2020}. 
Here, $\boldsymbol{\Sigma}^R$ and $\boldsymbol{\Psi}^R$ are row-covariance matrices with dimensions $P \times P$ and $F \times F$, and measure row-wise dependence of errors and disturbances, respectively.
Conversely, the matrices $\boldsymbol{\Sigma}^C$ and $\boldsymbol{\Psi}^C$ are column-covariance matrices with dimensions $Q \times Q$ and $G \times G$ that measure column-wise dependence of errors and disturbances, respectively. 
Dependent rows or columns are characterized by full covariance matrices, while independent row or columns are characterized by diagonal matrices  \citep{guptanagar}. Thus, the model is general enough to encopass various forms of dependence, while keeping the number of parameters low with respect to alternative full specifications of the covariance matrices. 

Since we deal with annual-based data describing the careers of different runners, we assume $\boldsymbol{Z} = \boldsymbol{I}_P$, 
$\boldsymbol{T} = \boldsymbol{I}_P$, $\boldsymbol{\Sigma} = \boldsymbol{I}_Q \otimes \boldsymbol{\Sigma}^R$ and $\boldsymbol{\Psi} = \boldsymbol{I}_G \otimes \boldsymbol{\Psi}^R$.
These assumptions imply that the states of different groups describing runners' performance across years are independent of each other and characterized by the same temporal dependence structures, which are those implied by local level models in which the trends of each discipline are discrete random walk \citep{durbinkoopman}. 
Assuming a priori that performance on discipline $p$ at time $t+1$ is a deviation from that at year $t$ seems reasonable, as a runner is not expected to progress or regress excessively from year to year. 
Further, setting $\boldsymbol{\Psi}^C$ to be diagonal implies that groups are independent of each other, a typical assumption in clustering. In this framework, the role of the states (i.e. $\boldsymbol{\alpha}_{p,t}^{(g)}$) is to describe various evolution of the performances of the runners. 
How these states evolve can be considered as the combination of many factors (e.g. individual traits, training, motivation, etc) that lead to unpredictable prior behaviors.
Further, conditional on the states and $\boldsymbol{S}$, there is no reason to assume the runners to be dependent between each other.
We note also that imposing $\boldsymbol{\Sigma}^C =\boldsymbol{I}_Q$  and  $\boldsymbol{\Psi}^C =\boldsymbol{I}_G$ are restrictions even stronger than required, but they help in stabilizing the estimation of the other components given the large number of missing observations present in the data.
Removing these restrictions is a delicate aspect in the predictive context in which we fit our model (see Section \ref{sec:posterior_inference}). Indeed, the aim of increasing flexibility struggles with the goal of reducing the variability of the estimates and predictions, due to both the large dimensions of the state space we are considering, but also by the need of adopting a diffuse prior specification and a large number of groups for well capturing the variability of the considered phenomena (see Section \ref{sec:prior_specification}).

\subsection{Missing data inform on clustering structure}
The previous section was developed conditional on all data being observed, i.e., when the runners run all $P$ distances during the entire period of observation.
However, this is not the case for data that describe the career trajectories of runners, since the lack of data is part of the career itself. 
To include these factors as informative aspect of runners' careers, we consider two other variables in the model. 
As first, we consider
\begin{align*}
d_{pq,t} = \begin{cases} 
1, \quad \text{if discipline $p$ for runner $q$ is observed at time $t$},
\\
0, \quad \text{otherwise},
\end{cases}
\end{align*}
to describe the presence or absence of the observed discipline for the runners.
Then we consider the variable $d_{q,t}^\star$ that informs whether the runner $q$ is in career during year $t$, which is
\begin{align*}
    d_{q,t}^\star  = 
    \begin{cases} 0, \quad \text{if runner $q$ has not started the career before $t$ (included),} \\
     1, \quad \text{if runner $q$ is in career during $t$}, \\
     2, \quad \text{if runner $q$ has finished the career in $t$ (included).}
    \end{cases}  
\end{align*}
The variable $d_{q,t}^\star$ is not decreasing in $t$, and describes the three possible states of runner's career.
Moreover, if $d_{q,t}^\star \in \{0,2\}$, then $d_{pq,t} = 0$ with probability $1$, for $ p =1,\ldots,P$, meaning that no distances are observed since the runner is not competing.
On the contrary, there might be runners with $d_{pq,t} = 0$, for $p = 1,\ldots, P$, even if $d_{q,t}^\star=1$. This is typical of runners who, despite being in a career, decide not to compete during one specific year, but compete in the following years. 

The division into three non-concurrent states allows for the introduction of temporal dynamics within the model of missing data patterns in an easy way. 
In particular, let $\mathbf{d}_{q}^\star = (d_{q,1}^\star,\ldots,d_{q,T}^\star)$,  $\mathbf{d}_{\cdot q,t} = (d_{1q,t},\ldots,d_{Pq,t})^\top$, and $\boldsymbol{D}_{q} = \begin{bmatrix}\mathbf{d}_{\cdot q,1} & \ldots & \mathbf{d}_{\cdot q,T}\end{bmatrix}$, $\mathcal{D} = \{\boldsymbol{D}_1,\ldots, \boldsymbol{D}_Q\}$, and $\mathcal{D}^\star = \{\mathbf{d}_{1}^\star, \ldots, \mathbf{d}_{Q}^\star\}$.
First, we make the following independence assumption among different subjects
\begin{align}
\label{indip2}
    p_{\boldsymbol{\theta}}(\mathcal{D}, \mathcal{D}^\star |\boldsymbol{S}) = \prod_{q=1}^Q     p_{\boldsymbol{\theta}} ( \boldsymbol{D}_{q}, \mathbf{d}_{q}^\star \vert S_q).
\end{align}
As a second step, we let $\mathbf{d}_{q}^\star$ and $\boldsymbol{D}_{q}$ be dependent on the group $S_q=g$ to which the runner $q$ belongs, and  make the following conditional independence assumption  
\begin{align}
\nonumber
    p_{\boldsymbol{\theta}} ( \boldsymbol{D}_{q}, \mathbf{d}_{q}^\star \vert S_q) &=      p_{\boldsymbol{\theta}} ( \boldsymbol{D}_{q} \vert  \mathbf{d}_{q}^\star, S_q)    p_{\boldsymbol{\theta}} ( \mathbf{d}_{q}^\star \vert S_q) \\
    \label{eq:conditioning_missing}&=
    \prod_{t=1}^T \bigg\{ \prod_{p=1}^P     p_{\boldsymbol{\theta}}( d_{pq,t} \vert d_{q,t}^\star, S_q ) \bigg\}    p_{\boldsymbol{\theta}}(d_{q,t}^\star \vert d_{q,t-1}^\star, S_q),
\end{align}
where $p_{\boldsymbol{\theta}}(d_{q,1}^\star \vert d_{q,0}^\star, S_q) = \lambda_{1g}^\star$ if $d_{q,1}^\star =1$, and $p_{\boldsymbol{\theta}}(d_{q,1}^\star \vert d_{q,0}^\star, S_q) =1-\lambda_{1g}^\star$ if $d_{q,1}^\star =0$.
Note that, in Equations \eqref{indip2} and \eqref{eq:conditioning_missing},  the subscript ${\boldsymbol{\theta}}$ in $p_{\boldsymbol{\theta}}( A \vert B) $  denotes conditional dependence of the form $p( A \vert B, \boldsymbol{\theta})$,  for slight abuse of notation, where $\boldsymbol{\theta}$ denotes a set of unknown parameter with finite dimensions (specified later).

In Equation \eqref{eq:conditioning_missing} we consider the following assumptions: for runner $q$, the conditional probabilities at time $t$ of transition from state $0$ to state $1$ is $p_{\boldsymbol{\theta}}(d_{q,t}^\star=1 \vert d_{q,t-1}^\star=0, S_q=g) = \lambda_{1g}^\star$ and from state $1$ to state $2$ is $p_{\boldsymbol{\theta}}(d_{q,t}^\star=2 \vert d_{q,t-1}^\star=1, S_q=g) = \lambda_{2g}^\star$. Both the probabilities are group dependent but constant over time.
By construction, the transitions from state $1$ to state $0$ or from state $2$ to states $0$ or $1$ are impossible events.
Further, for runner $q$, the conditional probability at time $t$ of observing a value for the generic discipline $p$ is $p_{\boldsymbol{\theta}}( d_{pq,t}=1 \vert d_{q,t}^\star=1, S_q =g)=\delta_{pg}$, which is group-dependent, but fixed over time.
Although transitions in the prevalence of the type of discipline done in a long career are possible for some runners (e.g. from shorter to longer distances), these transitions are difficult to detect with annual based data---which are summaries of the entire years---since it is enough to compete in only one race of the considered discipline to be included in the discipline-specific ranking lists.
Similarly, the assumption of constant probabilities during years used to describe the presence of missing values does not contemplate the possibility that runners would get seriously injured, and, thus, they would not compete in any discipline for more than a year.
Although there is no clear indication in the literature about the average duration and severity of an injury in middle distance runners \citep[see, e.g.,][]{van_gent_incidence_2007}, we assume here that severe injuries  (injuries that stop competitions for more than one year) are present only in low proportions, leaving open possible investigations on this aspect in the future.


\section{Likelihood and posterior distributions}
\label{sec:prior_specification}
\subsection{Likelihood and prior for the proposed model}
\noindent
Let $\mathcal{Y}$ denote the set of observations as if they were fully observed, $\mathcal{Y}^\star$  the set of variables which are effectively observed, and $\widetilde{\mathcal{Y}}$ the completion of $\mathcal{Y}^\star$, i.e. such that $\mathcal{Y} = \mathcal{Y}^\star \cup \widetilde{\mathcal{Y}}$ and $ \mathcal{Y}^\star 	\cap \widetilde{\mathcal{Y}} =$\O. 
Let also $\mathcal{A}=\{\boldsymbol{A}_1, \ldots, \boldsymbol{A}_T\}$ be the set storing the latent states of the state space model.
In order to derive the posterior distribution of the parameters, we present the likelihood of the observed process first, augmented for both the states $\mathcal{A}$, the missing observations $\widetilde{\mathcal{Y}}$, and $\boldsymbol{S}$.

The augmented likelihood is characterized by the following conditional independence structure
\begin{align}
\label{aug_likelihood_clus}
    p_{\boldsymbol{\theta}}(\mathcal{Y}, \mathcal{D}, \mathcal{D}^\star, \mathcal{A}, \boldsymbol{S}) = p_{\boldsymbol{\theta}} (\mathcal{Y} \vert \mathcal{D}, \mathcal{A}, \boldsymbol{S})     
    p_{\boldsymbol{\theta}}(\mathcal{D} \vert \mathcal{D}^\star, \boldsymbol{S})    p_{\boldsymbol{\theta}}( \mathcal{D}^\star \vert \boldsymbol{S}) p_{\boldsymbol{\theta}}(\boldsymbol{S}) p_{\boldsymbol{\theta}} (\mathcal{A}).
\end{align}
In Equation \eqref{aug_likelihood_clus},  $p_{\boldsymbol{\theta}} (\mathcal{Y} \vert \mathcal{D}, \mathcal{D}^\star, \mathcal{A}, \boldsymbol{S}) =  p_{\boldsymbol{\theta}} (\mathcal{Y} \vert \mathcal{D}, \mathcal{A}, \boldsymbol{S})$, and is determined by the measurement Equation \eqref{clust_meas_matr},  for which all observations are assumed to be available, and the prior on $\mathcal{A}$ is implicitly determined by the form of the state equation of the state space in Equation \eqref{clust_state_matr}. 
However, only $\mathcal{Y}^\star =\{\mathcal{Y}_1^\star,\ldots, \mathcal{Y}_T^\star\}$ is observed, but  $p_{\boldsymbol{\theta}} (\mathcal{Y} \vert \mathcal{D}, \mathcal{A}, \boldsymbol{S})$ can be obtained by conditioning, noting that
\begin{align*}
    p_{\boldsymbol{\theta}} (\mathcal{Y} \vert \mathcal{D}, \mathcal{A}, \boldsymbol{S}) =   p_{\boldsymbol{\theta}} (\mathcal{Y}^\star \vert \mathcal{D},  \mathcal{A}, \boldsymbol{S})   p_{\boldsymbol{\theta}} ( \widetilde{\mathcal{E}} \vert \mathcal{Y}^\star,  \mathcal{D}, \boldsymbol{S}),
\end{align*}
where $\widetilde{\mathcal{E}}$ stores all those entries in $\mathcal{E}= \{ \boldsymbol{E}_1,\ldots,\boldsymbol{E}_T\}$ associated with the missing values.
To  characterize $\boldsymbol{S}$, we make the following independence assumption \begin{align}
\label{indipdendence_assumption}
    p_{\boldsymbol{\theta}}(\boldsymbol{S}) = \prod_{q=1}^Q p_{\boldsymbol{\theta}}(\mathbf{s}_{q\cdot}) = \prod_{q=1}^Q \prod_{g=1}^G \pi_g^{\mathbbm{1}(S_q=g)},
\end{align}
where $\boldsymbol{\pi} = (\pi_g,\ldots,\pi_G)$ is such that $\pi_g \in (0,1)$, for $g=1,\ldots, G$, and $\sum_{g=1}^G \pi_g=1$.

As concerns model parameters, we assume that $\boldsymbol{\theta}$  factorizes as follows
\begin{align}
\label{prior_distribution}
    p(\boldsymbol{\theta}) = p(\widehat{\boldsymbol{A}}_{1\vert 0})  p(\boldsymbol{\Sigma}^R) p(\boldsymbol{\Psi}^R) p(\boldsymbol{\pi})   \prod_{g=1}^{G} \bigg\{p(\boldsymbol{\lambda}_g^\star)\prod_{p=1}^P p(\delta_{pg})\bigg\},
\end{align}
where $\boldsymbol{\lambda}_g^\star = (\lambda_{1g}^\star,\lambda_{2g}^\star)$. 
We further assume the probabilities driving missing data patterns to be uninformative Beta distributions, such that $\lambda_{1g}^\star \sim \text{Be}(1, 1)$, $\lambda_{2g}^\star \sim \text{Be}(1, 1)$, and $\delta_{pg} \sim \text{Be}(1, 1)$, for any $p=1,\ldots,P$ and $g=1,\ldots,G$. Covariance matrices are assumed to be Inverse Wishart distributions, such that $\mathbf{\Sigma}^R\sim \text{IW}_P(P+1, \boldsymbol{I}_P)$ and $\mathbf{\Psi}^R\sim \text{IW}_P(P+1, \boldsymbol{I}_P)$.
For what concerns the state space model, $\widehat{\boldsymbol{A}}_{1\vert 0}$ is assumed to follows a matrix-variate Normal distributions of mean $\bar{\mathbf{y}}_1 \mathbf{1}_G^\top$ and covariance $\boldsymbol{P}_{1\vert 0}= \boldsymbol{I}_G \otimes \boldsymbol{P}_{1\vert 0}^0$, with  $\boldsymbol{P}_{1\vert 0}^0=\text{diag}(p^2_1,\ldots,p^2_P)$,  where $\bar{\mathbf{y}}_1$ is the vector storing sample average of observed distances at first time instant and $p^2_p$ is twice the sample variance of the $p$–th observed discipline at the first time instant.
It is interesting to observe, however, that the number of parameters depends on the number of groups $G$, which is fixed. We consider an overfitting finite mixture specification of the model \citep[see, e.g.,][]{walli_sparse2016, walli_label}, in which $G$ is set to be large, and $\boldsymbol{\pi} = (\pi_1,\ldots,\pi_G) \sim \text{Dir}_G(e_1,\ldots, e_G)$ with hyper-parameters $e_1 = \ldots = e_G = 1/G$, the prior on the mixture weights favours emptying the extra components, leaving complete symmetry between the different components included in the model.
This assumption implies that, during the estimation procedure, the number of filled components may be lower than $G$, leading to the classical distinction between the number of clusters $G^{+}$ (i.e. the number of filled components) and the number of components $G$ included in the model, with $G^{+}\leq G$ \citep[see,][for an extensive discussion on the topic]{walli_sparse2016,walli_label,telescoping_bayesian}.
Under this prior specification, it is possible to derive a Gibbs sampling algorithm that involves all full conditionals that are conditionally conjugate. Its derivation is reported in the supplementary material. Draws of the states $\mathcal{A}$ are obtained using the simulation smoothing technique by \cite{sismo_koopman}, applied to a reduced form of the model derived using the reduction by transformation technique \citep[see][]{jungbacker2008likelihood}.
\subsection{Posterior distribution and alternative specifications}
The goal of our inferential procedure is to derive quantity of interests (e.g. predictive distributions) from a sample of the posterior distribution
\begin{align}
\label{posterior_distr}
    p^{\text{C}}(\boldsymbol{\theta}, \mathcal{A}, \boldsymbol{S}, \widetilde{\mathcal{E}} \vert \mathcal{Y}^\star, \mathcal{D}, \mathcal{D}^\star) \propto p(\boldsymbol{\theta}) p_{\boldsymbol{\theta}} (\mathcal{Y} \vert \mathcal{D}, \mathcal{A}, \boldsymbol{S})
    p_{\boldsymbol{\theta}}(\mathcal{D} \vert \mathcal{D}^\star, \boldsymbol{S})    p_{\boldsymbol{\theta}}( \mathcal{D}^\star \vert \boldsymbol{S}) p_{\boldsymbol{\theta}}(\boldsymbol{S}) p_{\boldsymbol{\theta}} (\mathcal{A}).
\end{align}
A sample from  the posterior distribution can be obtained using a Gibbs sampling scheme, as discussed in the supplementary material.  In Equation \eqref{posterior_distr}, the superscript ${}^\text{C}$ indicates that the considered posterior is referring to the \emph{complete} model (or model 1), because it assumes that both attitudes and history matter. 
By restricting the complete model in Equation \eqref{posterior_distr} it is possible to obtain a set of alternative reduced specifications, in which different missing data pattern schemes have different influence for clustering. 
More specifically, a set of alternative specifications can be derived by dropping the dependence on the selection matrix $\boldsymbol{S}$ in some parts of the model. We consider the following set of alternative specifications:
\begin{description}
\item[Model 2:] Missing data do not matters:
\begin{align}  
\label{model2}
p^{\text{NM}}(\boldsymbol{\theta}, \mathcal{A}, \boldsymbol{S}, \widetilde{\mathcal{E}} \vert \mathcal{Y}^\star, \mathcal{D}, \mathcal{D}^\star) \propto p(\boldsymbol{\theta}) p_{\boldsymbol{\theta}} (\mathcal{Y} \vert \mathcal{D}, \mathcal{A}, \boldsymbol{S})
         p_{\boldsymbol{\theta}}(\boldsymbol{S}) p_{\boldsymbol{\theta}} (\mathcal{A});
         \end{align}
\item[Model 3:] Only attitude matters:
\begin{align}  
\label{model3} p^{\text{A}}(\boldsymbol{\theta}, \mathcal{A}, \boldsymbol{S}, \widetilde{\mathcal{E}} \vert \mathcal{Y}^\star, \mathcal{D}, \mathcal{D}^\star) \propto p(\boldsymbol{\theta}) p_{\boldsymbol{\theta}} (\mathcal{Y} \vert \mathcal{D}, \mathcal{A}, \boldsymbol{S})
    p_{\boldsymbol{\theta}}(\mathcal{D} \vert \mathcal{D}^\star, \boldsymbol{S})   p_{\boldsymbol{\theta}}(\boldsymbol{S}) p_{\boldsymbol{\theta}} (\mathcal{A});\end{align}
\item[Model 4:] Only history matters:
\begin{align}\label{model4}
p^{\text{H}}(\boldsymbol{\theta}, \mathcal{A}, \boldsymbol{S}, \widetilde{\mathcal{E}} \vert \mathcal{Y}^\star, \mathcal{D}, \mathcal{D}^\star) \propto p(\boldsymbol{\theta}) p_{\boldsymbol{\theta}} (\mathcal{Y} \vert \mathcal{D}, \mathcal{A}, \boldsymbol{S})
        p_{\boldsymbol{\theta}}( \mathcal{D}^\star \vert \boldsymbol{S}) p_{\boldsymbol{\theta}}(\boldsymbol{S}) p_{\boldsymbol{\theta}} (\mathcal{A}).\end{align}
\end{description}
Alternative model specifications in Equations \eqref{model2}--\eqref{model4} have meaningful structural interpretations, if compared with the complete model in Equation \eqref{posterior_distr}.
In model $2$ both $p_{\boldsymbol{\theta}}(\mathcal{D} \vert \mathcal{D}^\star, \boldsymbol{S}) = p(\mathcal{D} \vert \mathcal{D}^\star)$ and $p_{\boldsymbol{\theta}}( \mathcal{D}^\star \vert \boldsymbol{S})=p( \mathcal{D}^\star)$, meaning that neither attitude nor history matter for clustering, and therefore are not correlated with the evolution of performances. 
In this case, missing data are still considered in the estimation procedure for obtaining $\widetilde{\mathcal{E}}$ and other elements related to the set of completed observations $\mathcal{Y}$ (e.g., $\boldsymbol{\Sigma}^R)$, but the part of likelihood describing the evolution of  $\mathcal{D}$ and $\mathcal{D}^\star$ is no longer dependent on $\boldsymbol{\theta}$ and $\mathbf{S}$.
In model $3$, instead, the attitude matters but history does not. 
This request is obtained by requiring $p_{\boldsymbol{\theta}}( \mathcal{D}^\star \vert \boldsymbol{S})= p( \mathcal{D}^\star)$. 
Note that the dependence of $p_{\boldsymbol{\theta}}(\mathcal{D} \vert \mathcal{D}^\star, \boldsymbol{S})$ on  $\mathcal{D}^\star$ is preserved, an important aspect because it is involved in the estimation of the parameters $\delta_{pg}$ related to runners' attitudes. 
Finally, in model $4$ $p_{\boldsymbol{\theta}}(\mathcal{D} \vert \mathcal{D}^\star, \boldsymbol{S}) = p(\mathcal{D} \vert \mathcal{D}^\star)$, leading to a model in which runners' attitudes do not matter for the evolution of the performances.
For simplicity, we do not distinguish $p(\boldsymbol{\theta})$ in the different models letting the elements included in $\boldsymbol{\theta}$ under different model specifications differ, based on the single case being considered (e.g. $\boldsymbol{\theta}=\{\hat{\boldsymbol{A}}_{1\vert 0}, \boldsymbol{\Sigma}^R, \boldsymbol{\Psi}^R, \boldsymbol{\pi}\}$  in model $2$). 

We can provide an interpretation to our model constructions from a two-step Bayesian learning perspective.
First, different structured prior on $\boldsymbol{S}$ are obtained, which simply reflect different clustering structures that we believe to be relevant for clustering the performances.
For example, for the complete model, this structured prior is
\begin{align*}
    p^{\text{C}}(\boldsymbol{\theta}, \boldsymbol{S} \vert  \mathcal{D}, \mathcal{D}^\star) \propto 
 p(\boldsymbol{\theta})
    p_{\boldsymbol{\theta}}(\mathcal{D} \vert \mathcal{D}^\star, \boldsymbol{S})    p_{\boldsymbol{\theta}}( \mathcal{D}^\star \vert \boldsymbol{S}) p_{\boldsymbol{\theta}}(\boldsymbol{S}).
\end{align*}
Second, the knowledge on the clustering structure is updated by considering the likelihood related to the performances, which is $p_{\boldsymbol{\theta}} (\mathcal{Y} \vert \mathcal{D}, \mathcal{A}, \boldsymbol{S}) p_{\boldsymbol{\theta}} (\mathcal{A})$.
Comparing different models allows to determine which of the four alternative specifications is most credible in explaining the observed variability in runners' performances.
Comparisons between models are obtained by assessing the models' abilities to predict the performance of out-of-sample runners, as explained in Section \ref{sec:competitors}.
 We note here that, in our model construction, performances depend directly on missing data patterns and that
the opposite direction, in which the performances have a direct influence on  the fact that runners remains (or not) in the sample, is not considered.
In general, it is easy to imagine that runners with unsatisfactory careers are more likely to leave competitions.
Our conjecture is that the decision of competing is a determinant factor for improving the performances, and this is motivated by both physiological and psychological considerations for which defining goals and training for achieving them can help in improving performances as well.
Further, although the priors on clustering structure does not depend on the performances, the posteriors do. 
The choice of considering a sufficiently large number of groups $G$ allows to account also for variability present in the data that may be effectively caused by cases in which performances have a direct impact on the choice of leaving the competitions. 
\section{Posterior inference and  out of sample predictions}
\label{sec:posterior_inference}
\subsection{Predictive inference}
Let $\mathcal{Y}_{[n]}^\star,$ $\boldsymbol{D}_{[n]}$, and $\mathbf{d}_{[n]}^\star$ denote 
the random variables describing, respectively, the performances, the participation in the distances and the history of a new runner $n$, not included in the sample. Let also $\mathbf{\Theta} = (\mathcal{A}, \boldsymbol{\theta})$ be the unknown elements which are shared across different runners, characterized by a posterior distribution $p^j(\mathbf{\Theta} \vert \mathcal{Y}^\star, \mathcal{D}, \mathcal{D}^\star)$ that can be obtained by means of an MCMC algorithm under model $j$.
We consider the following predictive density:
\begin{align}
    \label{predictive2}
    p^j(\mathcal{Y}_{[n]}^\star \vert  \boldsymbol{D}_{[n]}, \mathbf{d}_{[n]}^\star,  \mathcal{Y}^\star, \mathcal{D}, \mathcal{D}^\star) =  \int p^j(\mathbf{\Theta} \vert \mathcal{Y}^\star, \mathcal{D}, \mathcal{D}^\star) p^j_{\boldsymbol{\Theta}}(\mathcal{Y}^\star_{[n]} \vert \boldsymbol{D}_{[n]}, \mathbf{d}^\star_{[n]}) \text{d}\mathbf{\Theta},
\end{align}
for $j \in \{\text{C}, \text{NM}, \text{A}, \text{H}\}$.
In Equation \eqref{predictive2},  missing data patterns are supposed to be known and are treated as control variables that potentially have an influence on the predicted performances $\mathcal{Y}_{[n]}^\star$.
While the posterior distribution $p^j(\mathbf{\Theta} \vert \mathcal{Y}^\star, \mathcal{D}, \mathcal{D}^\star)$ is an output of the MCMC algorithms (see supplementary material),
the likelihood $p^j_{\boldsymbol{\Theta}}(\mathcal{Y}^\star_{[n]} \vert \boldsymbol{D}_{[n]}, \mathbf{d}^\star_{[n]})$ for the new individual $n$ can be obtained by marginalizing over groups as follows:
\begin{align*}
    p^j_{\boldsymbol{\Theta}}(\mathcal{Y}^\star_{[n]} \vert \boldsymbol{D}_{[n]}, \mathbf{d}^\star_{[n]}) = \sum_{g=1}^{G} p^j_{\boldsymbol{\Theta}}(S_{[n]}=g \vert \boldsymbol{D}_{[n]}, \mathbf{d}^\star_{[n]}) p^j_{\boldsymbol{\Theta}}(\mathcal{Y}^\star_{[n]} \vert \boldsymbol{D}_{[n]}, \mathbf{d}^\star_{[n]},  S_{[n]} = g).
\end{align*}
In the equation, the cluster allocation follows a Multinomial distribution characterized by weights
\begin{align*}
    p^j_{\boldsymbol{\Theta}}(S_{[n]}=g \vert \boldsymbol{D}_{[n]}, \mathbf{d}^\star_{[n]}) \propto p^j_{\boldsymbol{\Theta}}(S_{[n]}=g) p^j_{\boldsymbol{\Theta}}( \boldsymbol{D}_{[n]}, \mathbf{d}^\star_{[n]}\vert S_{[n]}=g),
\end{align*}
that depends on the likelihood related to cluster allocation $p_{\boldsymbol{\Theta}}(S_{[n]}=g)$  but also on the observed missing data patterns, that weigh differently the cluster allocation by means of $p_{\boldsymbol{\Theta}}( \boldsymbol{D}_{[n]}, \mathbf{d}^\star_{[n]}\vert  S_{[n]}=g)$. 
Next section details the use of the predictive distribution developed here for evaluating informativeness of missing data patterns under different model's specifications.
\subsection{Informativeness of missing data: out of sample comparison of alternative specifications}
\label{sec:competitors}
Let $\mathcal{Y}_{[1:N]}$ be a test set, storing the performances in different distances and  years of $N$ runners not included in the training sample for model estimation.
Let also $y_{p[n],t}$ be the generic scalar element denoting the performance of runner $n$ in discipline $p$ during year $t$. 
The quantity 
\begin{align*}
\mathbb{Q}^j(y_{p[n],t}) = p^j(y_{p[n],t} \vert  \boldsymbol{D}_{[n]}, \mathbf{d}_{[n]}^\star,  \mathcal{Y}^\star, \mathcal{D}, \mathcal{D}^\star),
\end{align*}
represents the predictive distribution obtained under model $j$, conditional on missing data patterns $(\boldsymbol{D}_{[n]}, \mathbf{d}_{[n]}^\star)$ and the set of available information $(\mathcal{Y}^\star, \mathcal{D}, \mathcal{D}^\star)$.
Let also $\widetilde{\mathbb{Q}}^j(y_{p[n],t})$ denote an approximation of $\mathbb{Q}^j(y_{p[n],t})$, given by a set of $B$ particles $\{y_{p[n],t}^{1},\ldots, y_{p[n],t}^{B}\}$.
It is possible to use the samples $\widetilde{\mathbb{Q}}^j(y_{p[n],t})$ to evaluate and compare the ability of our proposals in predicting the performances over different distances, for fixed missing data patterns described by $\boldsymbol{D}_{[n]}$, and $\mathbf{d}_{[n]}^\star$. 
We base our evaluations on the empirical counterpart of the \emph{continuous ranked probability score} (CRPS) and the \emph{interval score} \citep[see,][]{raftery2007,Kruger2021}, preferring models that minimize these scoring rules and that provide adequate prediction interval estimates in term of coverage and interval width.
The CRPS is defined as
\begin{align}
    S_1(\mathbb{Q}^j(y_{p[n],t})) = \int_{-\infty}^\infty\big\{\mathbb{Q}^j(y_{p[n],t})- \mathbbm{1}(y_{p[n],t}\leq z)\big\}^2 \text{d}z,
\end{align}
and the interval score is  defined as 
\begin{align*}
    S_2(\mathbb{Q}^j(y_{p[n],t})) = (u_\alpha^j-l_\alpha^j) + \frac{2}{\alpha} \big\{ (l_\alpha^j-y_{p[n],t})\mathbbm{1}(y_{p[n],t}<l_\alpha^j) + (y_{p[n],t}-u_\alpha^j)\mathbbm{1}(y_{p[n],t}>u_\alpha^j)\big\},
\end{align*}
where $l_\alpha^j$ and $u_\alpha^j$ are the $\alpha/2$ and $1-\alpha/2$ quantiles for the distribution  $\mathbb{Q}^j(y_{p[n],t})$, respectively, and $\alpha \in (0,1/2)$ is a fixed tolerance.
The interval score rewards narrow prediction intervals, and penalizes prediction intervals that do not include the observations. Details on these scores and their computation are reported in \cite{Kruger2021} (see, e.g., Equation 9).
It is relevant to highlight, however, that these scores are scale sensitive, and the scale of the predictions might depend both on the discipline $p$, the age $t$, as well as on the specific missing data patterns of runner $n$ we are considering.
For this reason, we propose to evaluate the predictions of a reference model $j$ and the prediction of an alternative model $j^\prime$ by means of the following score
\begin{align*}
    S_s^{j j^\prime}(\mathcal{Y}_{[1:N]}) = \frac{1}{\vert \mathcal{Y}_{[1:N]} \vert}\sum_{n=1}^{N} \sum_{y_{p[n],t} \in \mathcal{Y}_{[n]}}\mathbbm{1}(S_s(\mathbb{Q}^j(y_{p[n],t}) < S_s(\mathbb{Q}^{j^\prime}(y_{p[n],t})),
\end{align*}
for $s \in\{1,2\}$.
In the equation,  $\vert \mathcal{Y}_{[1:N]} \vert$ denotes the number of distinct observations present in the test set $\mathcal{Y}_{[1:N]}$. 
The scores  $S_s^{j j^\prime}(\mathcal{Y}_{[1:N]})$ range in $(0,1)$, and suggest that model $j$ is overall better than model $j^\prime$ if $S_s^{j j^\prime}(\mathcal{Y}_{[1:N]})>0.5$. 
\section{Real data analysis}
\label{sec:examples}
\subsection{Informativeness of missing data}
The models were estimated randomly splitting the runners into a training set, composed of $70\%$ of runners, and a test set with the remaining $30\%$.
Samples from the posterior distributions were obtained on the training set, using the last $2000$ of $10000$ iterations of the Gibbs sampling for each model. 
The number of components $G$ was fixed to $50$.
Of them, samples of size $2000$ from the predictive distributions in Equation \eqref{predictive2} were obtained, conditional on knowing the missing data patterns of runners in the test set (i.e. $\boldsymbol{D}_{[n]}$ and $\mathbf{d}_{[n]}^\star$, for $n = 1,\ldots, N$).

\begin{figure}
    \centering
    \includegraphics[scale = 0.32]{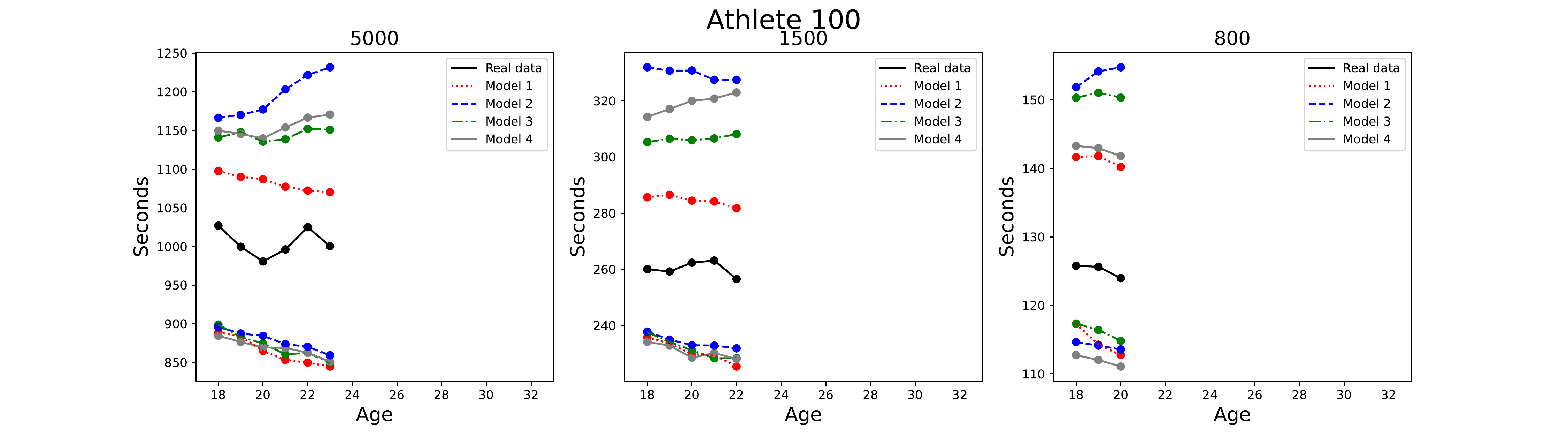}
    \includegraphics[scale = 0.32]{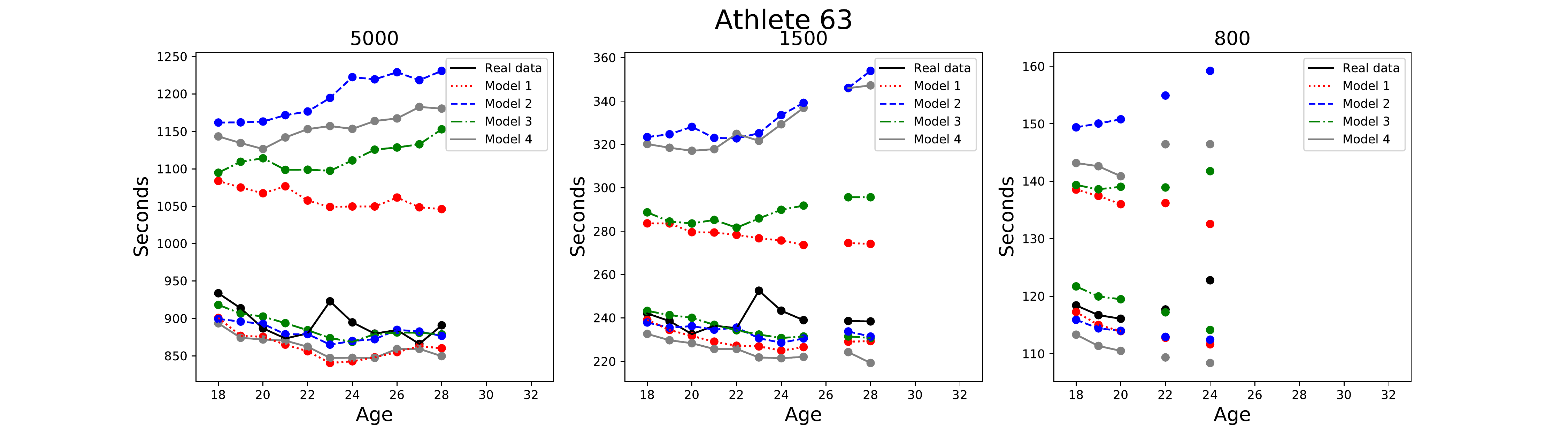}
    \includegraphics[scale = 0.32]{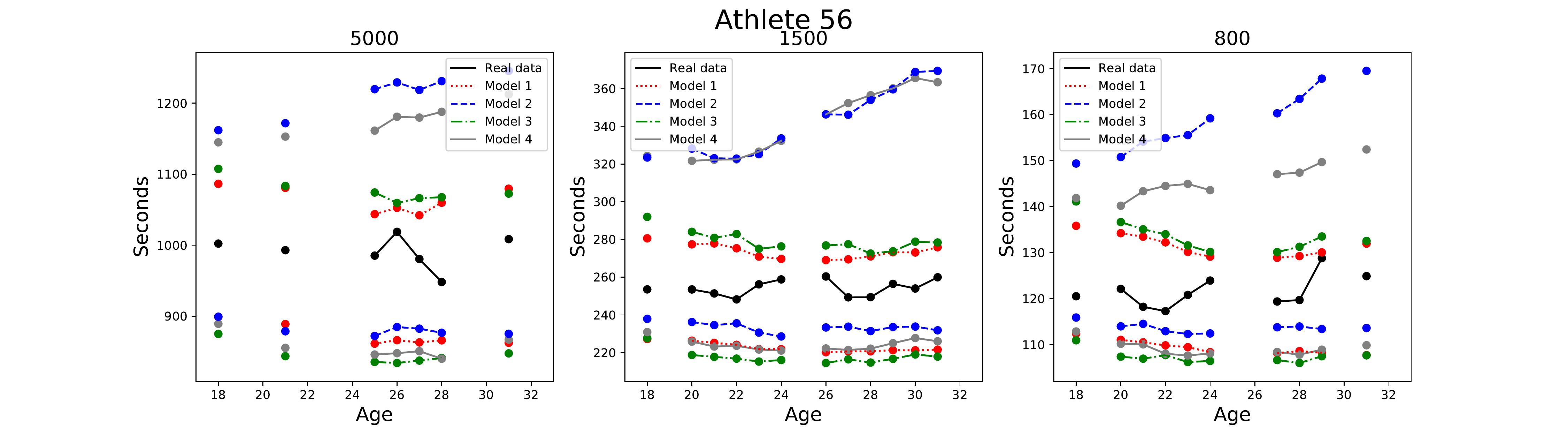}
    \includegraphics[scale = 0.32]{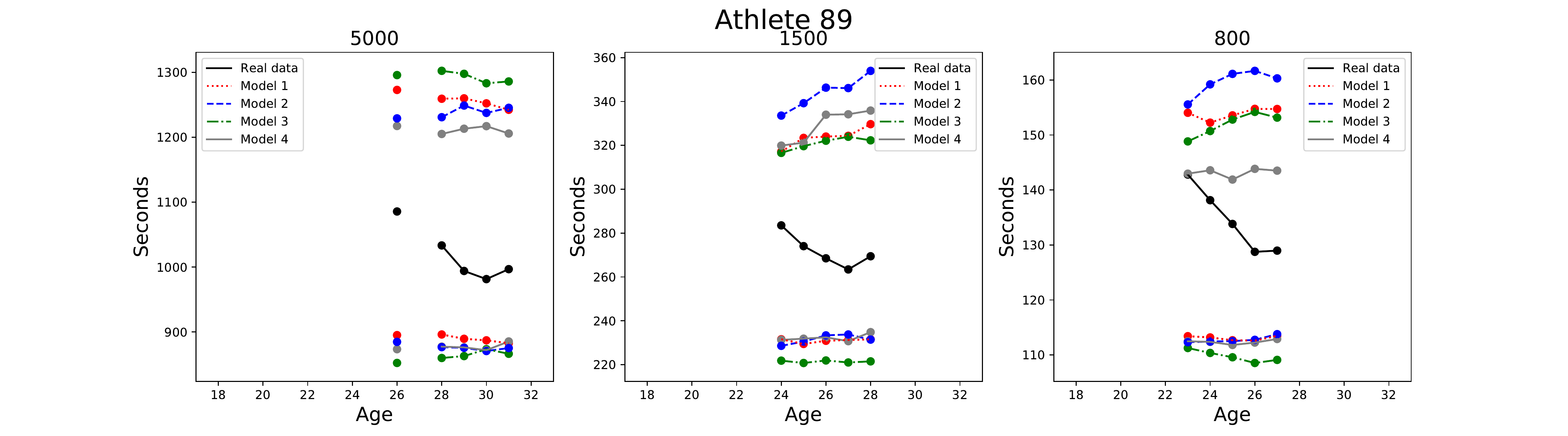}
    \caption{Quantile-based $95\%$ prediction intervals for the observed distances obtained conditional on knowing the missing data patterns of runners included in the test set. 
    The red-dotted lines represent the intervals for the model that treats missing data as informative,  while the blue-dashed lines represents the respective intervals for the model that does not treat them as such.
    The black lines represent the observed performance of the runners.}
    \label{fig:results1}
\end{figure}

The comparison over different models' predictions was done both graphically and using the scores described in Section \ref{sec:competitors}.
Results are summarized in Figure \ref{fig:results1} and Table \ref{tab:crps}. 
In the figures, the real data are represented by black solid lines, while the colored lines delimit $95\%$ quantile-based prediction bands.
A model is preferred if: (a) the real data lies within the prediction bands, (b) the predictions bands are narrower.
By looking at the results we can claim that model 1 (both attitude and history are considered as informative) and model 3 (only attitude is considered as informative) provide better results in terms of band widths, while including within the bands the real data.
As we note, different missing data patterns are represented in the figures.
We note the tendency of model 1 and model 3 of having lower upper limits of the band, an aspect that is even more evident for runners that participated to many competitions in different years (e.g. runners $100$,  $63$, and $56$).
Knowing that one runner has competed for many years reduces the uncertainty in the predictions by reducing the probability of observing bad performances. 
This highlights the effective presence of an association between the abilities of the runners in recording better performances with long histories and their participation in many competitions during the years.
Further, by comparing runner $89$ with runner $56$, for example, it is possible to grasp how entering later in competitions leads to more uncertainty in the performance predictions, increasing also the probabilities of observing worse performances. 
While the effect of knowing missing data patterns is clear when we compare the upper limits of the predictions bands, the movements of lower limits appear to be limited and less pronounced. 
This aspect is interesting because it points out that there are runners that, despite being characterized by short histories, are still able to perform satisfactorily when compared with those with longer histories.
These reasonings are conditional on the graphs shown, which are, of course, a selection of the runners in the test set.
The complete set of plots is reported in the supplementary material, showing a large number of runners with different missing data patterns and, as a consequence, different behaviors of the bands. 
\begin{table}
\caption{\label{tab:crps}Comparison between models given by $S_s^{j j^\prime}$ with the different scores. 
In this table, model $j$ (row) is compared to model $j^\prime$ (column) with respect the two scores.
Above the diagonal we report the scores for CRPS, below the diagonal we report the scores for IS, computed with $\alpha=0.05$. 
 A value above $0.5$ indicates preference of model $j$ with respect to model $j^\prime$. Remember that  $S_s^{j j^\prime}=1-S_s^{j^\prime j}$, for any score considered.}
\centering
\fbox{\begin{tabular}{l|rrrr}
 Model & Complete (1) & No missing (2) & Attitude (3) & History (4) \\ \hline
Complete (1) & -- & 0.574 & 0.537 & 0.520 \\
No missing (2) & 0.178 & -- & 0.402 & 0.395 \\
Attitude (3) & 0.364 & 0.757 & -- & 0.511 \\
History (4) & 0.233 & 0.680 & 0.306 & -- 
\end{tabular}}
\end{table}

The predictive scores computed with data of the test set suggest that the complete model is better than the others considering both the scores. 
However, the interval score suggest this aspect more markedly, highlighting the ability in outperforming the model that does not treat missing values as informative for around $80\%$ of the observations present in the test set.
For both the scores,  it results that the complete model is better than model $3$ (only attitude), which is better than model $4$ (only history), which is itself better than model $2$ (uninformative missing data patterns).
These results highlight how attitude and history (encopassed in the term missing data) seem to be effectively related to performances, giving support to the hypothesis that considering these aspect in the analysis of runners careers is definitely relevant and that, in this context, missing data have to be considered as informative.
Note that both the estimates of predictive scores are based on $741$ scalar observations which are characterized by different levels of dependence, so that a proper evaluation of uncertainty of these estimates is difficult.
In the supplementary material, a three-fold cross validation scheme shows how results seems to be stable. 
\subsection{Application}
\begin{figure}
    \centering
      {\includegraphics[scale=0.37]{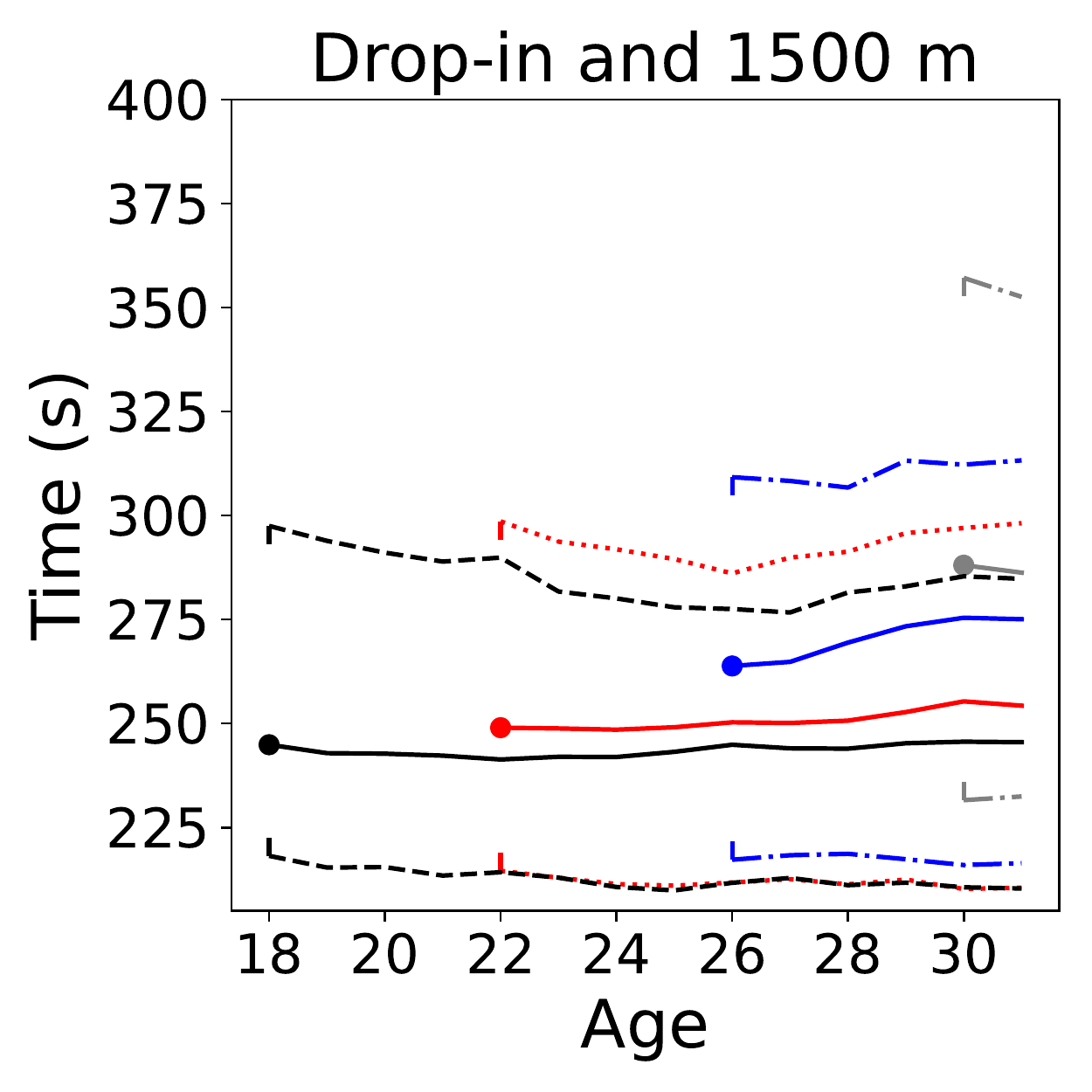}}
      {\includegraphics[scale=0.37]{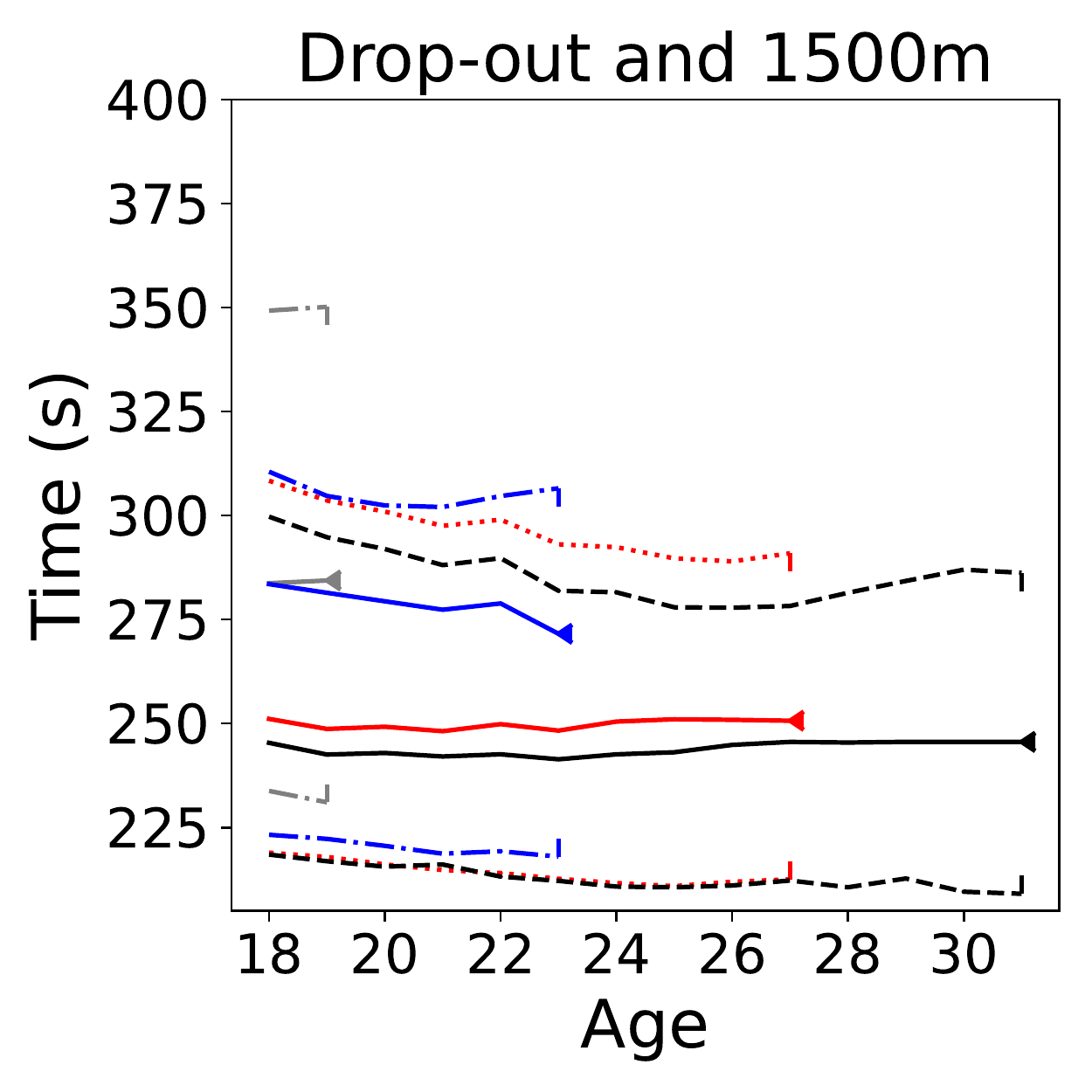}}\\
        \includegraphics[scale=0.37]{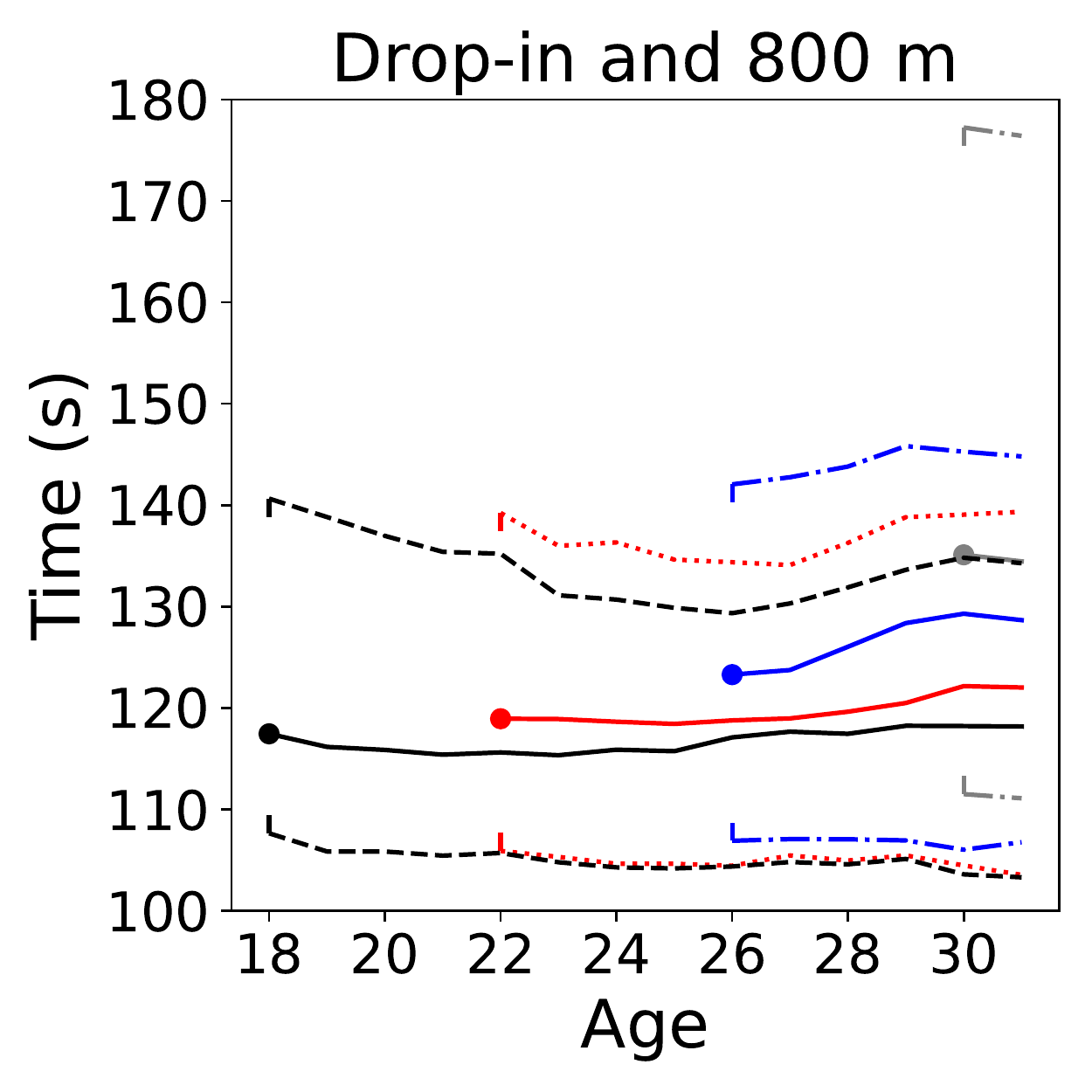}
    \includegraphics[scale=0.37]{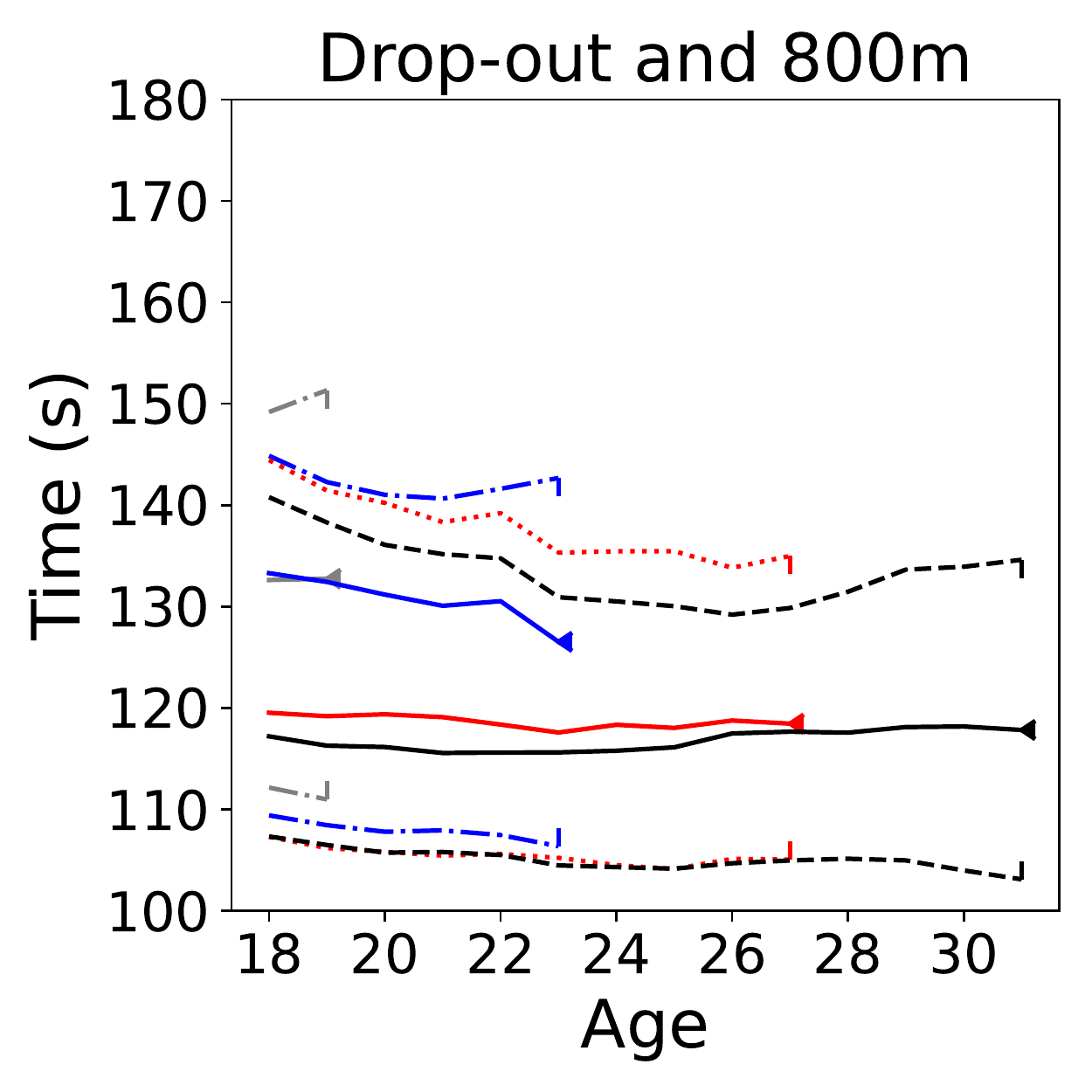}
    \caption{The association between drop-in and drop-out with performances in the $1500$ meters discipline, for an ideal runner that competes regularly in $1500$ (top) and $800$ meters (bottom). 
    On the left, the runner drops-in at ages $18$, $22$, $26$, $30$, respectively. On the right, the runner drops-out at age $20$, $24$, $28$ or after $32$ years old. Central solid line indicate the median of the predictive distributions. External lines indicate the $95\%$ confidence bands based on symmetric quantiles of the predictive distributions.}
    \label{fig:results_analysis}
\end{figure}

We illustrate how the complete model can be meaningful for sports scientists and coaches, answering to specific questions, based on a sample of size $2000$ of the predictive distributions explained in Section \ref{sec:posterior_inference}.
The first question is: how do late entry into competitions and early exit from competitions are related to performances?
To answer this question, we consider the conditional predictive distribution in Equation \eqref{predictive2}, in which we vary the age at which the runner enters or exits from the sample, letting the runner participate in both $800$ and $1500$ discipline for all the years of his career.
Comparing the different distributions of the performance allows to catch how the uncertainty related to the predicted performance changes, according to the different histories considered. 
Figure \ref{fig:results_analysis} show the results of our procedure for the $1500$ and $800$ meters distances.
Solid central lines represent the median of the predictive distributions over different ages. Quantile-based $95\%$ prediction bands are on the contrary represented with different dashed lines, that denote the respective lower an the upper limits. 
For what concerns drop-in, we note from the left panels that late entry into competition is associated with worse median performances over the years, with lower limits of the predictive confidence bands that worsen only for runners that entry into competition at ages $26$ and $30$.
The upper limits, on the contrary, seem to rise with later drop-in. 
Based on these results, we can say that for the ideal runners we are considering,  later entry in career can still permit to reach good levels, but it is much more likely that their performances will be worse with respect to runners with a longer career (earlier drop-in).
A similar reasoning applies for drop-out, in the right panels. 
The median of the predictive distributions seems indeed to be worse for runners that drop-out earlier in their life, with the upper limits that seem to show more variation with respect to the lower. 
In this case, it is not unlikely to expect runners that drop-out earlier, although their performances seems good, but it is more unlikely that runners that compete for more years record worse results. 
Similar results for different scenarios can be obtained with an analogous approach and are reported in the supplementary material.
\par
\begin{figure}
    \centering
    {\includegraphics[scale=0.37]{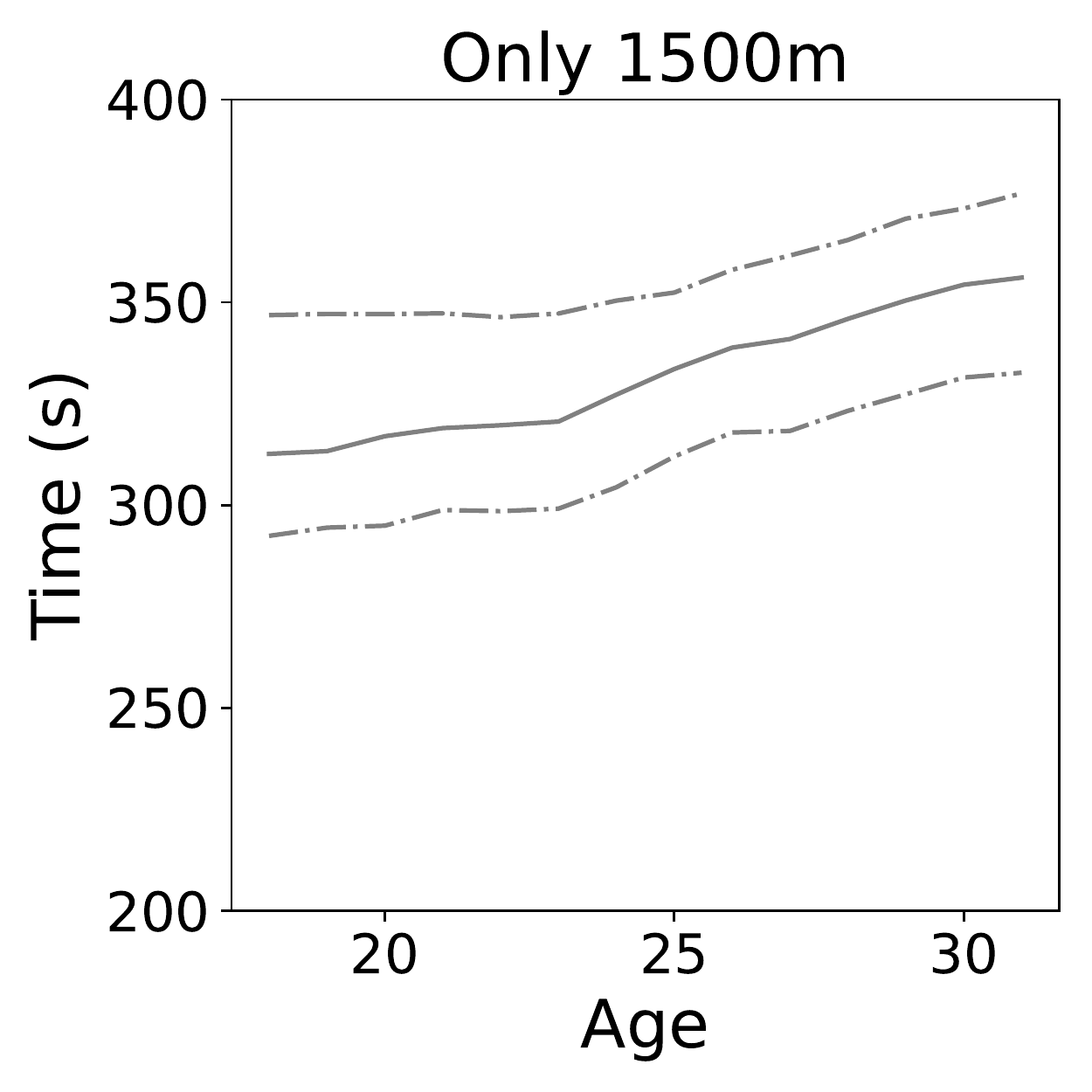}}
    {\includegraphics[scale=0.37]{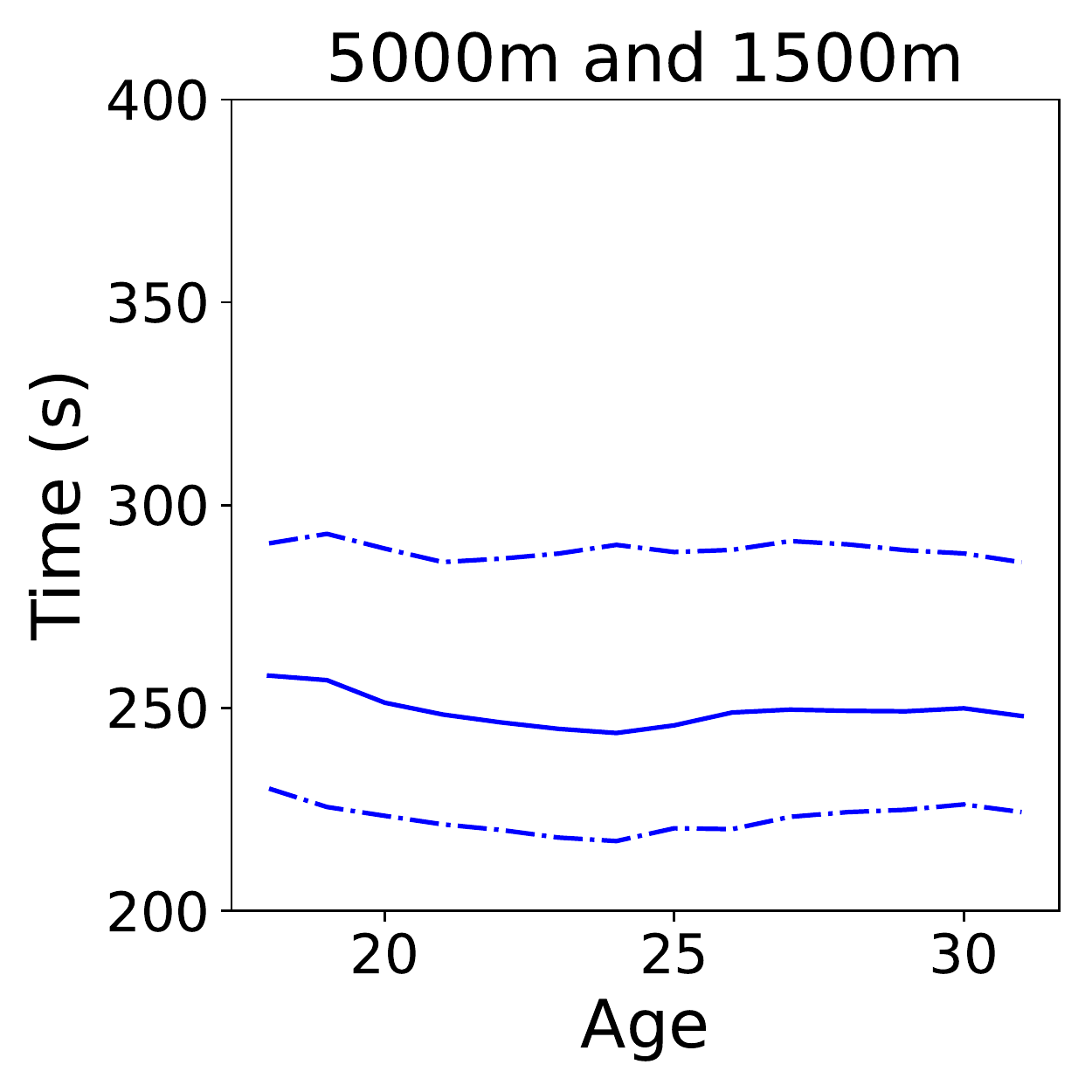}}\\
   {\includegraphics[scale=0.37]{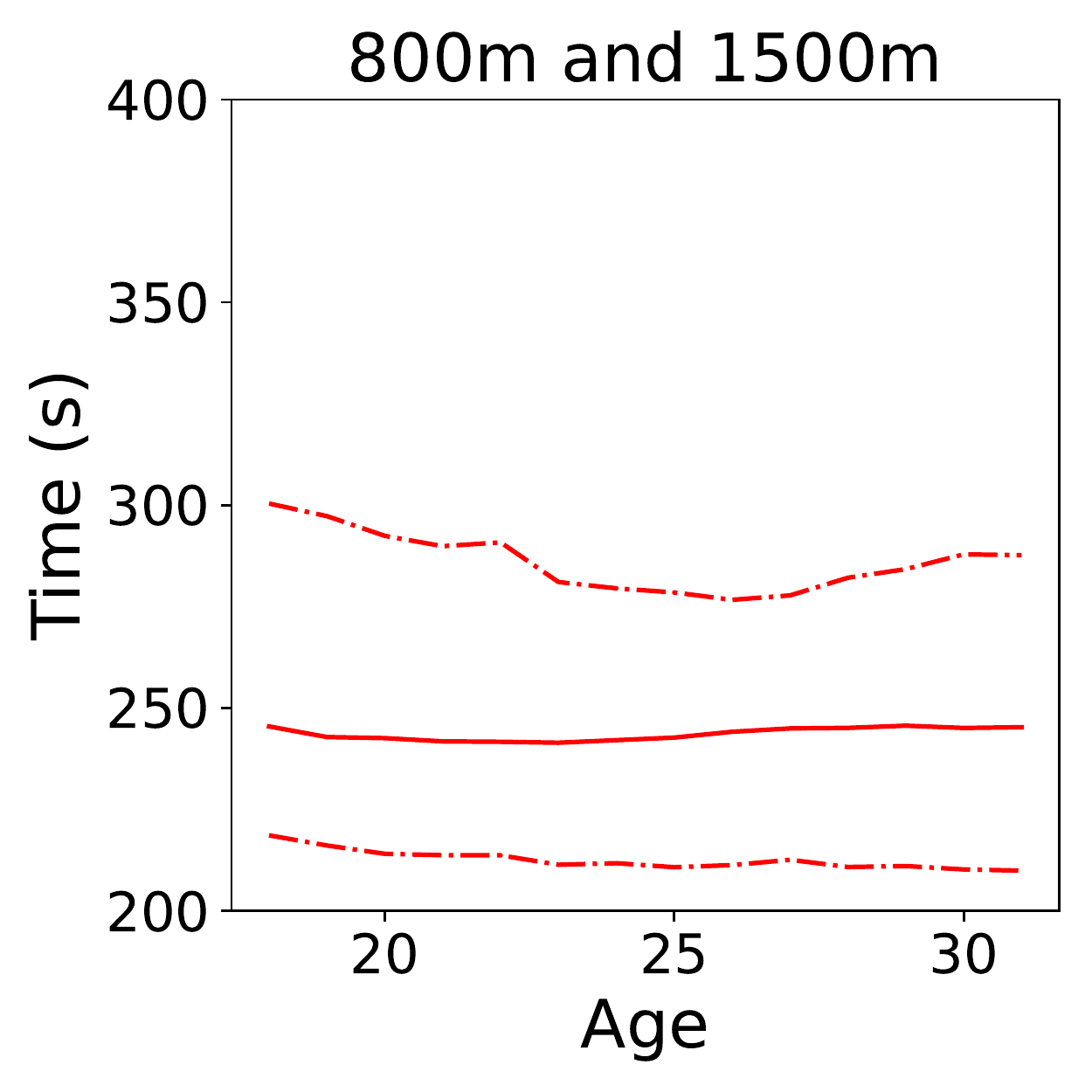}}
    {\includegraphics[scale=0.37]{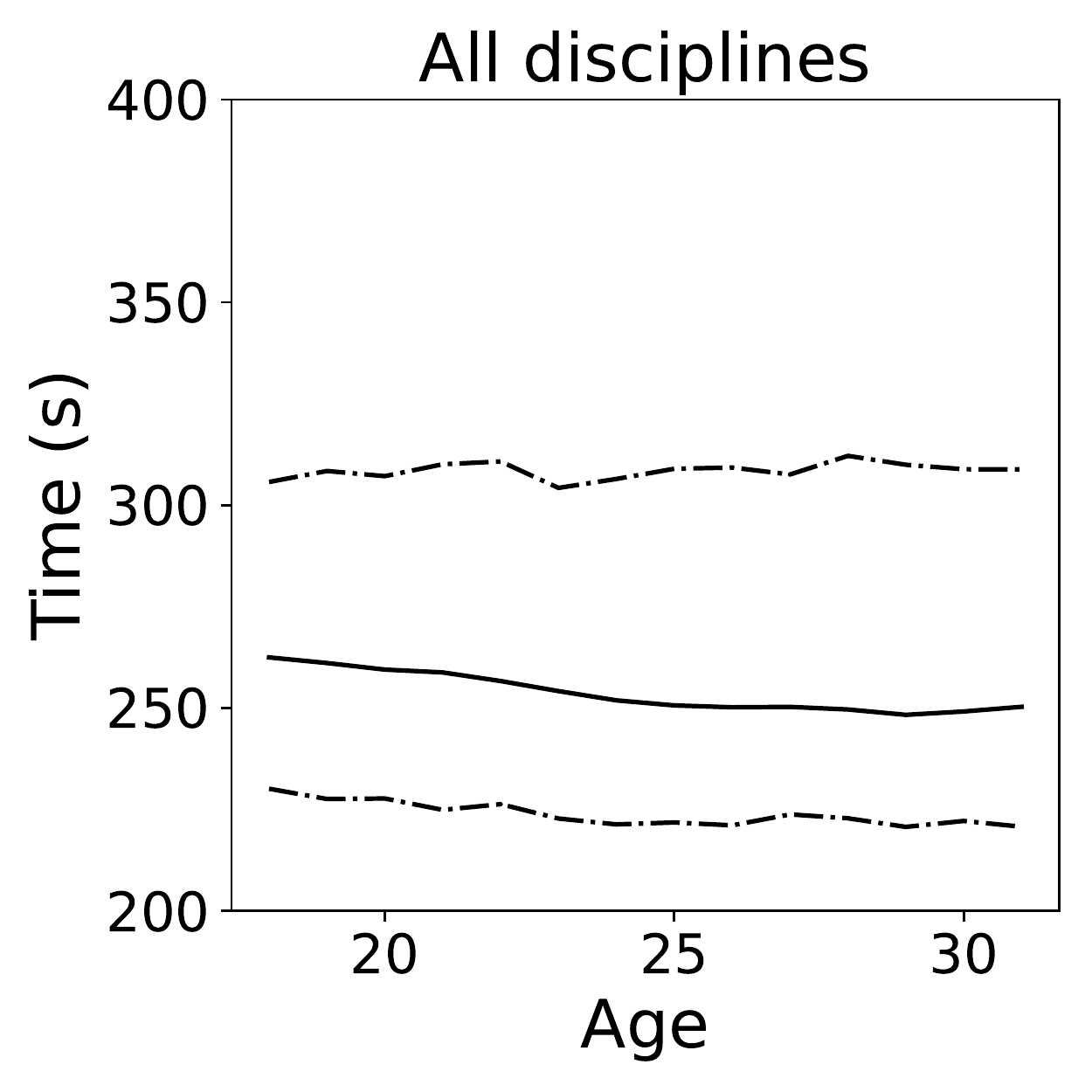}}
    \caption{Predictive distributions of performances in the $1500$ meters discipline, obtained for four different runners with different ways of participation in other distances during the years. }
    \label{fig:attitudes}
\end{figure}

\indent
The second question is: how does competing in different disciplines impact the performances?
In this case, we consider runners with a complete observed career, and that compete, every year, in the distances with different strategy:
the first runner competes only in the $1500$ meters discipline; the second runner competes in $1500$m and $5000$m distances; the third one competes in $1500$m and $800$m distances; the fourth runner competes in all distances. 
Results are shown in Figure \ref{fig:attitudes}, in terms of their respective predictive distributions in $1500$ meters discipline. By comparing the results, we see clearly how, in the training sample, worse performances appear to be associated with the choice of competing only in one discipline over the years.
On the contrary, being a runner that competes in more than one discipline seems to be associated with better performances, with differences between the respective predictive distributions which are less evident. More specifically, the greater differences can be seen comparing runners that compete in two distances with the one that competes in all three.
Indeed, both the limits of the bands and the median of the predictive distribution appear to be shifted up for the runner that competes in all distances, implying a slightly worse performance for these kind of runners.
Based on our results, competing in all distances seems not to be  an optimal choice to achieve better results in the $1500$ meters discipline. On the contrary, runners that specialize in $1500$ and $5000$ meters or in $1500$ and $800$ meters distances seem to have better performances overall, especially for the latter type of runners.
Further analyses with other distances are left in the supplementary material.
\section{Conclusions}
We have investigated whether prediction of the runners' performance is improved by an accurate assessment of the presence of missing data patterns. 
Our analysis has provided strong evidence that for our data, missing data patterns are informative in predicting performances and they constitute a structural part of the signal explaining the observed variability of the runners' performances. 

The statistical analysis took place via a matrix-variate state space model, in which the observed trends were clustered by employing a selection matrix involved in the measurement equation, and by storing the unknown cluster allocations of the runners. To include observed missing data patterns as informative on the clustering structure, two distinct processes were included in the model.
The first included  the runner's history as potentially informative by considering when an runner starts or stops competing. The second aimed to include as potentially informative the runner's attitude by considering in which distances the runner mostly participates.

Our results based on out of sample comparisons suggest that it is important to consider both these processes when describing runners' performances whereas considering only attitude is better than considering only history.
There is evidence for a deterioration in performance when one runner starts competing later or finishes earlier
and for improvement for runners that compete regularly in more distances compared to runners that compete only in one distance.  Finally, competing in all three distances does not seem to be associated to better performances with respect to competing in two adjacent distances, such as $800$ and $1500$ meters or $1500$ and $5000$ meters.

Our key message is to illustrate the usefulness of considering missing data when describing runners' performances.
Our modelling framework can be immediately used in data with alternative distances, such as sprinting, throwing, or multidisciplinary distances such as of heptathlon and decathlon. It would be also interesting to investigate whether these findings differ in female runners or in countries with possibly different coaching methodologies. 

\section*{Acknowledgment}This research was supported by funding from the University of Padova Research Grant 2019-2020, under grant agreement BIRD203991.
\section*{Supplementary material} Please write to the authors for supplementary material, including code, data, proofs and derivations.

\bibliographystyle{rss}
\bibliography{references}
\end{document}